\documentclass[%
 reprint,
 amsmath,amssymb,
pra,
]{revtex4-1}

\usepackage{graphicx}% Include figure files
\usepackage{setspace}
\usepackage{dcolumn}% Align table columns on decimal point
\usepackage{minibox}
\usepackage{bm}% bold math
\usepackage{hyperref}% add hypertext capabilities
\usepackage{float}%positions the figures
\usepackage{tikzfig}
\tikzstyle{every picture}=[baseline=-0.25em]
\tikzstyle{dotpic}=[scale=0.6]
\tikzstyle{diredges}=[every to/.style={diredge}]
\tikzstyle{dot graph}=[shorten <=-0.1mm,shorten >=-0.1mm,scale=0.6]
\tikzstyle{plot point}=[circle,fill=black,minimum width=2mm,inner sep=0]

\tikzstyle{braceedge}=[decorate,decoration={brace,amplitude=2mm,raise=-1mm}]
\tikzstyle{small braceedge}=[decorate,decoration={brace,amplitude=1mm,raise=-1mm}]
\tikzstyle{left hook arrow}=[left hook-latex]
\tikzstyle{right hook arrow}=[right hook-latex]

\tikzstyle{dot}=[inner sep=0.7mm,minimum width=0pt,minimum height=0pt,fill=black,draw=black,shape=circle]
\tikzstyle{white dot}=[dot,fill=white]
\tikzstyle{alt white dot}=[white dot,label={[xshift=2.9mm,yshift=-0.1mm]left:$\cdot$}]
\tikzstyle{gray dot}=[dot,fill=gray!50]
\tikzstyle{box vertex}=[draw=black,rectangle]
\tikzstyle{whitebg}=[fill=white,inner sep=2pt]
\tikzstyle{graph state vertex}=[sg vertex,fill=black]

\tikzstyle{wide point}=[fill=white,draw=black,shape=isosceles triangle,shape border rotate=90,isosceles triangle stretches=true,inner sep=1pt,minimum width=1.5cm,minimum height=5mm]
\tikzstyle{wide copoint}=[fill=white,draw=black,shape=isosceles triangle,shape border rotate=-90,isosceles triangle stretches=true,inner sep=1pt,minimum width=1.5cm,minimum height=5mm]
\tikzstyle{symm}=[ultra thick,shorten <=-1mm,shorten >=-1mm]

\tikzstyle{square box}=[rectangle,fill=white,draw=black,minimum height=6mm,minimum width=6mm]
\tikzstyle{square gray box}=[rectangle,fill=gray!30,draw=black,minimum height=6mm,minimum width=6mm]
\tikzstyle{point}=[regular polygon,regular polygon sides=3,draw=black,scale=0.75,inner sep=-0.5pt,minimum width=7mm,fill=white]
\tikzstyle{copoint}=[point,regular polygon rotate=180,fill=white]
\tikzstyle{gray point}=[point,fill=gray!40!white]
\tikzstyle{gray copoint}=[copoint,fill=gray!40!white]

\newcommand{\edgearrow}{{\arrow[black]{>}}}
\newcommand{\edgetick}{{\arrow[black,scale=0.7,very thick]{|}}}
\tikzstyle{diredge}=[postaction=decorate,decoration={markings, mark=at position 0.55 with \edgearrow}]
\tikzstyle{medium diredge}=[postaction=decorate,decoration={markings, mark=at position 0.75 with \edgearrow}]
\tikzstyle{short diredge}=[->]
\tikzstyle{halfedge}=[-)]
\tikzstyle{other halfedge}=[(-]
\tikzstyle{freeedge}=[(-)]
\tikzstyle{white edge}=[line width=5pt,white]
\tikzstyle{tick}=[postaction=decorate,decoration={markings, mark=at position 0.5 with \edgetick}]
\tikzstyle{small map edge}=[|-latex, gray!60!blue, shorten <=0.9mm, shorten >=0.5mm]
\tikzstyle{thick dashed edge}=[very thick,dashed,gray!40]
\tikzstyle{map edge}=[|-latex,very thick, gray!40, shorten <=1mm, shorten >=0.5mm]
\tikzstyle{tickedge}=[postaction=decorate,
  decoration={markings, mark=at position 0.5 with \edgetick}]
\tikzstyle{dirtickedge}=[postaction=decorate,
  decoration={markings, mark=at position 0.5 with \edgetick},
  decoration={markings, mark=at position 0.85 with \edgearrow}]
\tikzstyle{dirdoubletickedge}=[postaction=decorate,
  decoration={markings, mark=at position 0.4 with \edgetick},
  decoration={markings, mark=at position 0.6 with \edgetick},
  decoration={markings, mark=at position 0.85 with \edgearrow}]

\tikzstyle{arrs}=[-latex,font=\small,auto]
\tikzstyle{arrow plain}=[arrs]
\tikzstyle{arrow dashed}=[dashed,arrs]
\tikzstyle{arrow bold}=[very thick,arrs]
\tikzstyle{arrow hide}=[draw=white!0,-]
\tikzstyle{arrow reverse}=[latex-]
\tikzstyle{cdnode}=[]

\tikzstyle{cnot}=[fill=white,shape=circle,inner sep=-1.4pt]
\tikzstyle{wire label}=[font=\footnotesize, auto]

\floatstyle{boxed}

\usepackage{xy}
\xyoption{matrix}
\xyoption{frame}
\xyoption{arrow}
\xyoption{arc}

\usepackage{ifpdf}
\ifpdf
\else
\PackageWarningNoLine{Qcircuit}{Qcircuit is loading in Postscript mode.  The Xy-pic options ps and dvips will be loaded.  If you wish to use other Postscript drivers for Xy-pic, you must modify the code in Qcircuit.tex}
\xyoption{ps}
\xyoption{dvips}
\fi

\entrymodifiers={!C\entrybox}

\newcommand{\qw}[1][-1]{\ar @{-} [0,#1]}
\newcommand{\qwx}[1][-1]{\ar @{-} [#1,0]}
\newcommand{\gate}[1]{*+<.6em>{#1} \POS ="i","i"+UR;"i"+UL **\dir{-};"i"+DL **\dir{-};"i"+DR **\dir{-};"i"+UR **\dir{-},"i" \qw}
\newcommand{\measure}[1]{*+[F-:<.9em>]{#1} \qw}
\newcommand{\measureD}[1]{*{\xy*+=<0em,.1em>{#1}="e";"e"+UR+<0em,.25em>;"e"+UL+<-.5em,.25em> **\dir{-};"e"+DL+<-.5em,-.25em> **\dir{-};"e"+DR+<0em,-.25em> **\dir{-};{"e"+UR+<0em,.25em>\ellipse^{}};"e"+C:,+(0,1)*{} \endxy} \qw}
\newcommand{\control}{*!<0em,.025em>-=-<.2em>{\bullet}}
\newcommand{\ctrl}[1]{\control \qwx[#1] \qw}
\newcommand{\targ}{*+<.02em,.02em>{\xy ="i","i"-<.39em,0em>;"i"+<.39em,0em> **\dir{-}, "i"-<0em,.39em>;"i"+<0em,.39em> **\dir{-},"i"*\xycircle<.4em>{} \endxy} \qw}
    \newcommand{\targo}[1]{*+<.02em,.02em>{\xy ="i","i"-<.39em,0em>;"i"+<.39em,0em> **\dir{-}, "i"-<0em,.39em>;"i"+<0em,.39em> **\dir{-},"i"*\xycircle<.4em>{} \endxy}  \qwx[#1] \qw}
\newcommand{\qswap}{*=<0em>{\times} \qw}
\newcommand{\push}[1]{*{#1}}
\newcommand{\gategroup}[6]{\POS"#1,#2"."#3,#2"."#1,#4"."#3,#4"!C*+<#5>\frm{#6}}
\newcommand{\rstick}[1]{*!L!<-.5em,0em>=<0em>{#1}}
\newcommand{\lstick}[1]{*!R!<.5em,0em>=<0em>{#1}}
\newcommand{\Qcircuit}{\xymatrix @*=<0em>}
\newcommand{\ket}[1]{\ensuremath{\left|#1\right\rangle}}

\usepackage{graphicx}% Include figure files
\usepackage{dcolumn}% Align table columns on decimal point
\usepackage{bm}% bold math
\pgfdeclarelayer{edgelayer}
\pgfdeclarelayer{nodelayer}
\pgfsetlayers{edgelayer,nodelayer,main}

\tikzstyle{none}=[inner sep=0pt]
\definecolor{hexcolor0xff0000}{rgb}{1.000,0.000,0.000}
\definecolor{hexcolor0x000000}{rgb}{0.000,0.000,0.000}
\definecolor{hexcolor0xffff00}{rgb}{1.000,1.000,0.000}
\definecolor{Lime}{rgb}{0.000,1.000,0.000}
\definecolor{hexcolor0x030000}{rgb}{0.012,0.000,0.000}

\tikzstyle{rn}=[circle,fill=hexcolor0xff0000,draw=hexcolor0x000000,line width=0.8 pt, minimum width=8pt, minimum height=8pt, inner sep=0.4pt]
\tikzstyle{gn}=[circle,fill=Lime,draw=hexcolor0x000000,line width=0.8 pt, minimum width=8pt, minimum height=8pt, inner sep=0.4pt]
\tikzstyle{yn}=[circle,fill=hexcolor0xffff00,draw=hexcolor0x000000,line width=0.8 pt, minimum width=8pt, minimum height=8pt, inner sep=0.4pt]
\tikzstyle{diam}=[shape=diamond,fill=hexcolor0x030000,draw=hexcolor0x000000,minimum width=8pt, minimum height=8pt, inner sep=0pt]
\tikzstyle{Had}=[rectangle,fill=hexcolor0xffff00,draw=hexcolor0x000000,minimum width=8pt, minimum height=8pt, inner sep=0.2pt]

\tikzstyle{simple}=[-,draw=hexcolor0x000000,line width=1.000]
\tikzstyle{arrow}=[-,draw=hexcolor0x000000,postaction={decorate},decoration={markings,mark=at position .5 with {\arrow{>}}},line width=2.000]
\tikzstyle{tick}=[-,draw=hexcolor0x000000,postaction={decorate},decoration={markings,mark=at position .5 with {\draw (0,-0.1) -- (0,0.1);}},line width=2.000]

\begin{document}

\preprint{APS/123-QED}

\title{Complete set of circuit equations for Stabilizer Quantum Mechanics}% Force line breaks with \\

\author{Andr$\acute{e}$ Ranchin$^{1,2}$}
\author{Bob Coecke$^{1}$}
 
\affiliation{
$^{1}$University of Oxford, Department of Computer Science, Quantum Group\\
$^{2}$Imperial College London, Department of Physics, Controlled Quantum Dynamics
}

\date{\today}% It is always \today, today,

\begin{abstract}

We find a sufficient set of equations between quantum circuits from which we can derive any other equation between stabilizer quantum circuits. To establish this result, we rely upon existing work on the completeness of the graphical ZX language for quantum processes. The complexity of the circuit equations, as opposed to the very intuitive reading of the much smaller number of ZX-equations, advocates the latter for performing computations with quantum circuits.  

\end{abstract}

\maketitle

\section{Introduction}

Studying quantum theory from a computational and information-theoretic point of view has provided important no-go theorems \cite{Be64,Ko67,Wo82,Bar96,Pu11}, a description of new physical phenomena \cite{Be92,Be93,Ben92,Zi05} and a better understanding of the importance of quantum resources, like entanglement \cite{Ho09}. The development of quantum computation as a sub-discipline of computer science in its own right, moreover, leads us to ask important new questions that would not normally occur to physicists. 

There are several natural logical properties that are important with regard to quantum computation. The first one of these is universality. It has been shown \cite{Deu95,Ba95} that a universal set of gates for any quantum computation consists of single qubit gates and the controlled-not gate. This means that any valid quantum circuit can be built up using composition and tensor products of gates in this universal set.

Two other essential properties for logical systems are soundness and completeness. Previous work has focused on whether abstract diagrammatical systems are sound or complete for quantum mechanics \cite{Sel12,Du08,Back12}. Here, we wish to present soundness and completeness in a more concrete setting by describing them in terms of familiar quantum circuits. This should clarify the importance of these logical properties from the viewpoint of quantum computation, in analogy with the work done on the role of universality in quantum computation.

Assume that we are given a set of equations between quantum circuits. New circuit equations can be obtained by locally substituting parts of circuits by equal quantum circuits.

\textit{Soundness} guarantees that any equation between quantum circuits that can be deduced from an original set of equations is in agreement with quantum theory. A set of circuit equations is sound if each quantum circuit equation in the original set of equations agrees with quantum mechanics and if any equation built from this original set is also in agreement with quantum theory.

\textit{Completeness} ensures that any equation between quantum circuits that is true in quantum theory can be deduced from the original set of equations. A complete set of circuit equations for quantum mechanics is one from which the equality of any two quantum circuits corresponding to the same physical process can be deduced. Although constructing a set of circuit equations that is sound for quantum theory is simple, finding a complete set of circuit equations is far from trivial. Such a set, if it exists, would provide a logical set of axioms from which one could formally derive whether or not any two quantum processes are equivalent. 

In this article, we restrict the search for a complete set of circuit equations to a subclass of quantum mechanics, namely stabilizer quantum theory. A stabilizer quantum mechanics process consists of tensor products and compositions of computational basis state preparations, Clifford unitaries and measurements of observables in the Pauli group (or at least one of these three). Two such physical processes are equivalent if they can be described by exactly the same quantum circuit. 

This naturally leads us to ask the following question:\\

\textit{Can one find a sound and complete set of quantum circuit equations from which one can deduce the equivalence of any two stabilizer processes?} \\

We answer this question in the affirmative. The crux of the proof draws from converting an abstract graphical calculus into quantum circuits. 

In the following, we construct a logical circuit calculus whose elements correspond to physical stabilizer processes. We show that this calculus is equivalent to an abstract graphical calculus called the ZX network \cite{Du08}. 

This demonstrates that familiar quantum circuits can always be used instead of the algebraic calculus to study stabilizer theory. However, since the ZX network diagrams are not restricted to the structure of circuits, the ZX network is a more flexible and convenient tool for calculation. The abstract calculus relies on reasoning with diagram elements which have no explicit physical interpretation. 

The elements of the circuit calculus, on the other hand, correspond directly to physical systems and processes. Therefore, we can use this graphical language to study the physical theory of stabilizer quantum mechanics from a logical point of view. This allows us to explicitly present a \textbf{complete set of quantum circuit equations for stabilizer quantum mechanics}. 

This is an important result towards understanding the logic of stabilizer quantum mechanics: this complete set of circuit equations is a set of axioms from which any two stabilizer quantum circuits which are identical can be proven to be the same. Note that the existence of such a definable complete set of \textit{circuit equations} cannot be deduced from only studying the abstract ZX network. 

\section{Stabilizer quantum theory}
A very useful subclass of quantum mechanical operations is stabilizer quantum mechanics. 
Stabilizer states are eigenstates with eigenvalue 1 of each operator in a subgroup of the Pauli group:
 
\hspace{30pt} $P_n:= \{ \alpha g_1 \otimes ... \otimes g_n : \alpha \in \{\pm1, \pm i\}$, with 

\hspace{30pt}  $g_k \in \{I, \sigma_x, \sigma_y, \sigma_z \}, \forall k \in \{1, ..., n \} \}$. \\
The Clifford group is the group of unitary operations: 
\begin{center}
$C_n := \{ U : U g U^{\dagger} \in P_n, \forall g \in P_n \}$. \\
\end{center}
It is generated by the phase, Hadamard and C-NOT gates. \\
Stabilizer quantum mechanics \cite{Got97} includes state preparations in the computational basis, Clifford unitaries and measurements of observables in the Pauli group. This non-universal subclass of quantum mechanics is particularly important for a large number of quantum protocols, including quantum teleportation\cite{Be93}, super-dense coding \cite{Be92} and quantum key distribution \cite{Gis02}. It also underlies the current theory of quantum error correction.

By the Gottesman-Knill theorem\cite{GotKn98}, stabilizer quantum mechanics can be efficiently simulated by a classical computer. 
It has been shown \cite{Ed11} \cite{Puse12} that there is a close relationship between the stabilizer formalism and Spekkens' toy theory \cite{Spek07}. 

Independently from work in this paper, a recent result \cite{Sel13} presents a rewrite system by which any Clifford operator can be reduced to a unique normal form.

\section{ZX network}

We will now describe the ZX network \cite{Du08,Du11}, which is a two-colored pictorial calculus aiming to reproduce certain aspects of quantum theory. This calculus directly allowed us to find the complete set of circuit equations for stabilizer quantum mechanics presented below.

\begin{figure}[H]
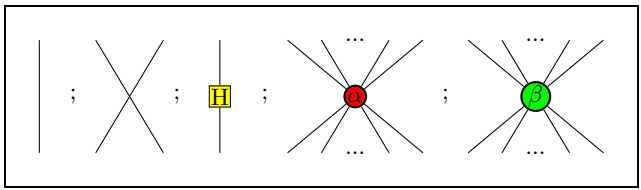

\begin{center}
\frame{
}
\end{center}
\caption{Generating diagrams for the ZX network.}
\label{ZXgen}
\end{figure}

\begin{figure}[ht!]
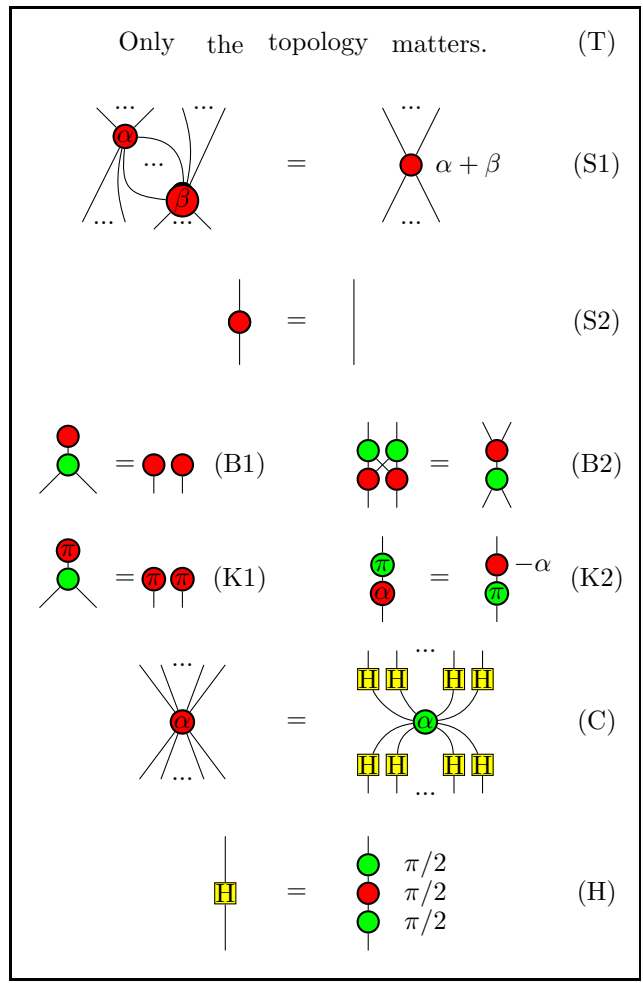

\frame{%
}
\caption{Diagrammatic rules for the ZX network.}
\label{ZXRules}
\end{figure}

General network diagrams are built out of parallel (tensor product) and downward compositions of generating diagrams from Figure \ref{ZXgen}. 

The axioms of the ZX network are summarized in Figure \ref{ZXRules}. The (T) rule means that after identifying the inputs and outputs of any part of a ZX network, any topological deformation of the internal structure does not matter. The (H) rule was introduced in \cite{Dun09}.

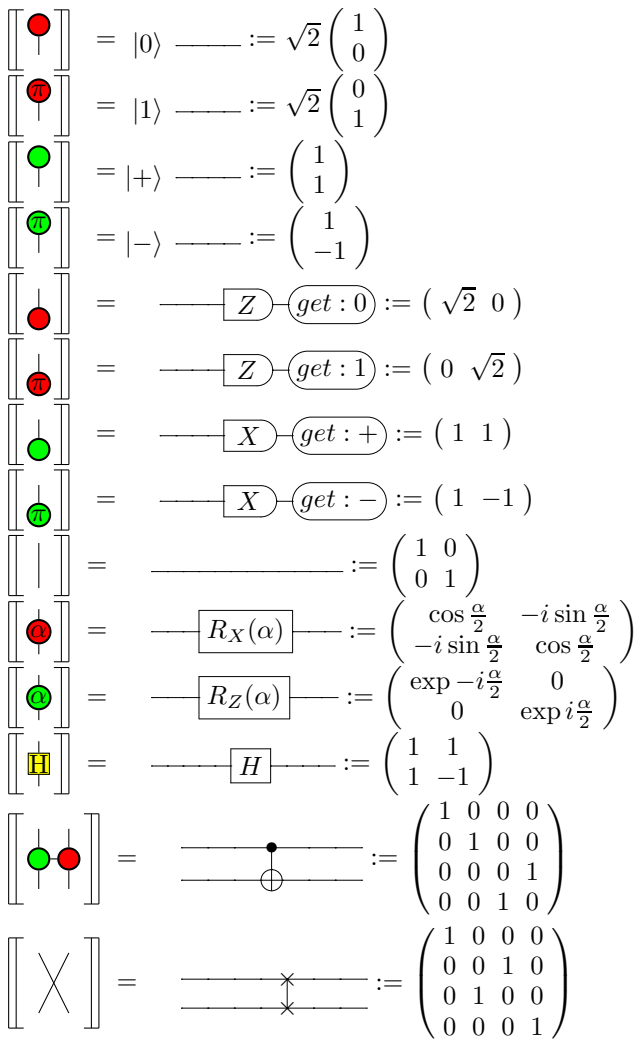
\begin{figure}[ht]
\begin{flushleft}
	 \minibox{ \hspace{3pt} = \hspace{16pt}} \minibox{\Qcircuit @C=.6em @R=.5em {
\lstick{\ket{0}}&\qw &\qw   &\qw   & \qw  }} := $\sqrt{2} \left( \begin{array}{c} 1 \\ 0
\end{array}%
 \right)$
\end{flushleft}
\caption{Quantum circuit interpretation of the ZX network elements.}
		\label{Circ}
\end{figure}
 
Two network diagrams can be shown to be equal by locally replacing some part of a diagram with a diagram equal to it. 

ZX network diagrams are logical elements which have no explicit physical meaning and can be modeled in many different ways. A particular interpretation in terms of quantum circuits can be constructed from the diagrams of the ZX network as shown in Figure \ref{Circ}. The ZX network is universal for quantum computation since any quantum circuit can be built in this way.

We know that the ZX network is sound for quantum mechanics: if two diagrams are equal according to the rules of the ZX network then their corresponding quantum circuits are equivalent \cite{Du08}. Note that the converse is not true: it can be impossible, from the axioms, to show the equality of two ZX network diagrams whose corresponding quantum circuits are equivalent. The ZX network simplifies numerous quantum calculations. It allows us to study a number of fundamental aspects of  quantum theory from a high-level mathematical point of view \cite{Dun10,Hors11,Coe10}. \\

\textbf{Theorem (Backens)}\cite{Back12}: The ZX network is complete for stabilizer quantum mechanics. \\

This means that any equation between two ZX network diagrams (put into matrix mechanics) which can be shown to be true using stabilizer quantum mechanics is derivable using the rules of the ZX network. Note that this completeness result only requires the axioms in Figure \ref{ZXRules} to hold with phases $\alpha$ and $\beta$ in the set $\{ -\pi/2, 0, \pi/2, \pi \}$.  

\section{Quantum circuits for the ZX network axioms}

This section presents the formal proof of the result stated in the introduction.   

In light of Backens' theorem, the quantum circuit equations corresponding to the axioms of the ZX network will be complete for stabilizer quantum mechanics. First of all, note that directly using Figure \ref{Circ} to convert the ZX network axioms into equations between linear operators does not yield a complete set of equations between quantum circuits since some of the resulting equations between linear operators cannot be expressed as quantum circuit equalities.

Therefore, in order to obtain the desired set of sound and complete circuit equations for stabilizer theory, we need to clarify the relationship between the ZX network and quantum stabilizer circuits. In order to do this formally, we introduce a symmetric monoidal category of stabilizer quantum circuits and show that it is equivalent to the symmetric monoidal category of the ZX network: \\   

\textbf{Lemma:} There is an equivalence of categories between the free symmetric monoidal categories of quantum circuits $\mathcal{F}_{SMC}(Circ)$ and of the ZX network $\mathcal{F}_{SMC}(ZX)$ (quotient to their axioms):
\begin{center}
$\mathcal{F}_{SMC}(Circ)/\equiv_{Circ} \leftrightarrow \mathcal{F}_{SMC}(ZX) / \equiv_{ZX}$.
\end{center}
$\mathcal{F}_{SMC}(Circ)$ is a free symmetric monoidal category over the monoidal signature \cite{Kiss12}: 

$\mathcal{S}:=\{CNOT;SWAP;prepare\ket{0};prepare\ket{+};$ \\
\hspace{35pt}  $postselect\ket{0},postselect\ket{+}, R_x(\alpha);R_z(\beta) \}$.

These are the consistuent `gates' of the symmetric monoidal category, which can be combined using composition and the tensor product. \\

The axioms for the category $\mathcal{F}_{SMC}(Circ)$, which are quantum circuit equations corresponding directly to the axioms of the ZX network ($\mathcal{F}_{SMC}(ZX)$), are given in Figure \ref{stabcirc}. This gives us a new insight into the structure of the ZX network, namely an understanding of what the axioms of the network mean, in terms of familiar quantum circuits. 

This equivalence of categories means that there exists a full, faithful, essentially surjective functor $[[\cdot]]: \mathcal{F}_{SMC}(ZX)/\equiv_{ZX}  \rightarrow \mathcal{F}_{SMC}(Circ)/\equiv_{Circ}$. For the constructive proof of the existence of this functor, we use the functor $[[\cdot]]$ in Figure \ref{Circ} and check that it is full, faithful and essentially surjective. \\

In practice, this requires us to find a set of ZX network equations which are equivalent to the axioms of the ZX network ($\equiv_{ZX}$) and are in a form that can be directly related to quantum circuits using Figure \ref{Circ}. Such a set of ZX network circuit-like equations is shown in Figure \ref{Circax}, in Appendix A. If we use the quantum circuit equations obtained by applying the functor in Figure \ref{Circ} to the network equations in Figure \ref{Circax} as the axioms $\equiv_{Circ}$ for the category $\mathcal{F}_{SMC}(Circ)$, then $[[\cdot]]: \mathcal{F}_{SMC}(ZX)/\equiv_{ZX}  \rightarrow \mathcal{F}_{SMC}(Circ)/\equiv_{Circ}$ is full, faithful and essentially surjective by construction. \\

Appendix A proves that the set of equations in Figure \ref{Circax} are equivalent to the ZX network axioms. These ZX network equations can be directly related to the axioms $\equiv_{Circ}$ for the category $\mathcal{F}_{SMC}(Circ)$ in Figure \ref{stabcirc}, using the functor in Figure \ref{Circ}. Note that the equivalence in this lemma holds for arbitrary phases $\alpha$ and $\beta$ in the ZX network axioms.    

\section{A complete set of circuit equations for stabilizer quantum mechanics}

The Lemma from the previous section shows that any quantum circuit equation which, when written in the ZX network, can be shown to be true using the ZX axioms from Figure \ref{ZXRules}, can be shown to be true using the equivalent circuit equations in Figure \ref{stabcirc}.
 
Backens' theorem states that any quantum circuit equation which can be shown to be true using stabilizer quantum mechanics is derivable using the ZX axioms when written as an equation between two ZX network diagrams. 

Combining the Lemma with the fact that the ZX network is sound for stabilizer quantum mechanics shows that any equation between quantum circuits which can be derived from the circuit equations in Figure \ref{stabcirc} is in agreement with stabilizer quantum mechanics.

Synthesizing these results yields the main result of this paper:
   
\textbf{Theorem:} The set of quantum circuit equations in Figure \ref{stabcirc} with phases $\alpha$ and $\beta$ in the set $\{ -\pi/2, 0, \pi/2, \pi \}$ is both sound and complete for stabilizer quantum mechanics. \\

\begin{widetext}

We now present the sound and complete set of quantum circuit equations for stabilizer quantum mechanics: \\

\minibox{\Qcircuit @C=.6em @R=.5em {
\lstick{\ket{0}} &\qw  & \qw & \qw & \qw & \targ  & \qw & \qw   \\
\lstick{\ket{0}} &\qw  & \qw &  \targ  & \qw & \ctrl{-1} & \qw & \qw    \\
&\qw & \qw &\ctrl{-1} & \qw &  \qw    & \qw & \qw   
}} \minibox{ \hspace{10pt} \large = \hspace{20pt} }
\minibox{\Qcircuit @C=.6em @R=.5em {
&\qw & \qw & \ctrl{1} & \qw & \qw & \qw & \qw   \\
\lstick{\ket{0}} &\qw  & \qw &  \targ  & \qw & \ctrl{1} & \qw & \qw    \\
\lstick{\ket{0}} &\qw   & \qw  & \qw  & \qw & \targ & \qw & \qw   
}} \minibox{ \hspace{70pt}\Qcircuit @C=.6em @R=.5em {
\lstick{\ket{+}} &\qw  & \qw & \qw & \qw & \ctrl{1}  & \qw & \qw   \\
\lstick{\ket{+}} &\qw  & \qw & \ctrl{1}   & \qw & \targ & \qw & \qw    \\
&\qw & \qw & \targ & \qw &  \qw    & \qw  & \qw  
}}
\minibox{ \hspace{10pt} \large = \hspace{20pt}}
\minibox{\Qcircuit @C=.6em @R=.5em {
&\qw & \qw & \targ  & \qw & \qw & \qw & \qw   \\
\lstick{\ket{+}} &\qw  & \qw & \ctrl{-1}  & \qw & \targ & \qw  & \qw   \\
\lstick{\ket{+}} &\qw   & \qw  & \qw  & \qw & \ctrl{-1} & \qw  & \qw  
}}\begin{flushright}
\minibox{\hspace{30pt} \large (S1circ)}
\end{flushright}

\scalebox{0.85}{\minibox{\Qcircuit @C=.6em @R=.5em {
 &\qw  & \targ & \qw & \qw & \qw  & \measureD{X} & \measure{ get: +} \\
 &\qw  & \ctrl{-1} &  \qw  & \targ & \qw & \measureD{X} & \measure{ get: +}  \\
&\qw & \qw & \qw & \ctrl{-1} &  \qw & \qw     
}}} \minibox{ \large = } 
\scalebox{0.85}{\minibox{\Qcircuit @C=0.6em @R=.5em {
&\qw & \qw & \qw & \ctrl{1} & \qw & \qw  \\
 &\qw  & \ctrl{1} &  \qw  & \targ  & \qw &  \measureD{X} & \measure{ get: +}    \\
 &\qw   & \targ  & \qw  & \qw & \qw &  \measureD{X} & \measure{ get: +} 
}}} \hspace{10pt} \scalebox{0.85}{\minibox{\hspace{10pt}\Qcircuit @C=.6em @R=.5em {
 &\qw  & \ctrl{1} & \qw & \qw & \qw  & \measureD{Z} & \measure{ get: 0} \\
 &\qw  & \targ &  \qw  & \ctrl{1} & \qw & \measureD{Z} & \measure{ get: 0} \\
&\qw & \qw & \qw & \targ &  \qw & \qw     
}}} \minibox{  \large = } 
 \scalebox{0.85}{\minibox{ \Qcircuit @C=.6em @R=.5em {
&\qw & \qw & \qw & \targ & \qw & \qw \\
 &\qw  & \targ &  \qw  & \ctrl{-1}  & \qw & \measureD{Z} & \measure{ get: 0}  \\
 &\qw   & \ctrl{-1}  & \qw  & \qw & \qw & \measureD{Z} & \measure{ get: 0}
}}} \\[0.8in]

\scalebox{0.85}{\minibox{\Qcircuit @C=.6em @R=.5em {
 &\qw  & \qw & \qw & \qw & \targ  & \qw & \measureD{Z} & \measure{ get: 0}  \\
\lstick{\ket{0}} &\qw  & \qw &  \targ  & \qw & \ctrl{-1} & \qw & \qw  \\
&\qw & \qw &\ctrl{-1} & \qw &  \qw    & \qw & \qw
}}} \minibox{ \large =  }
\scalebox{0.85}{\minibox{\Qcircuit @C=.6em @R=.5em {
&\qw & \qw & \targ & \qw & \qw & \qw & \measureD{Z} & \measure{ get: 0}   \\
&\qw  & \qw &  \ctrl{-1}  & \qw & \ctrl{1} & \qw & \qw   \\
\lstick{\ket{0}} &\qw   & \qw  & \qw  & \qw & \targ & \qw & \qw  
}}} \hspace{25pt} \scalebox{0.85}{\minibox{ \hspace{2pt}\Qcircuit @C=.6em @R=.5em {
 &\qw  & \qw & \qw & \qw & \ctrl{1}  & \qw & \measureD{X} & \measure{ get: +}  \\
\lstick{\ket{+}} &\qw  & \qw & \ctrl{1}   & \qw & \targ & \qw & \qw  \\
&\qw & \qw & \targ & \qw &  \qw    & \qw  & \qw
}}}\minibox{ \large = }\scalebox{0.85}{\minibox{\Qcircuit @C=.6em @R=.5em {
&\qw & \qw & \ctrl{1}  & \qw & \qw & \qw & \measureD{X} & \measure{ get: +} \\
 &\qw  & \qw & \targ  & \qw & \targ & \qw & \qw  \\
\lstick{\ket{+}} &\qw   & \qw  & \qw  & \qw & \ctrl{-1} & \qw  & \qw
}}}\begin{flushright}
\minibox{ \large (S2circ)}
\end{flushright}

\scalebox{0.85}{\minibox{\Qcircuit @C=.6em @R=.5em {
\lstick{\ket{+}} &\qw  & \targ & \qw & \qw & \qw  & \qw \\
 &\qw  & \ctrl{-1} &  \qw  & \targ & \qw & \measureD{X} & \measure{ get: +}  \\
&\qw & \qw & \qw & \ctrl{-1} &  \qw & \qw     
}}} \minibox{  \large = \hspace{2pt}} \scalebox{0.85}{\minibox{\Qcircuit @C=.6em @R=.5em {
 \lstick{\ket{+}} &\qw & \qw & \qw & \targ & \qw & \qw  \\
 &\qw  & \ctrl{1} &  \qw  & \ctrl{-1}  & \qw &  \qw    \\
 &\qw   & \targ  & \qw  & \qw & \qw &  \measureD{X} & \measure{ get: +} 
}}} \hspace{25pt} \scalebox{0.85}{\minibox{ \hspace{10pt}\Qcircuit @C=.6em @R=.5em {
\lstick{\ket{0}} &\qw  & \ctrl{1} & \qw & \qw & \qw  &  \qw\\
 &\qw  & \targ &  \qw  & \ctrl{1} & \qw & \measureD{Z} & \measure{ get: 0} \\
&\qw & \qw & \qw & \targ &  \qw & \qw     
}}} \minibox{ \large = \hspace{2pt}} 
 \scalebox{0.85}{\minibox{ \Qcircuit @C=.6em @R=.5em {
\lstick{\ket{0}} &\qw & \qw & \qw & \ctrl{1} & \qw & \qw \\
 &\qw  & \targ &  \qw  & \targ  & \qw & \qw  \\
 &\qw   & \ctrl{-1}  & \qw  & \qw & \qw & \measureD{Z} & \measure{ get: 0}
}}} \\[0.7in]

\begin{center}
\minibox{\Qcircuit @C=.6em @R=.5em {
\lstick{\ket{0}} &\qw &\qw & \targ & \qw & \qw   \\
&\qw  & \qw &  \ctrl{-1} & \qw  & \qw    
}} \minibox{ \hspace{10pt} \large = \hspace{20pt} }
\minibox{\Qcircuit @C=.6em @R=.5em {
\lstick{\ket{0}}&\qw &\qw  & \targ & \qw & \qswap & \qw & \qw    \\
&\qw  & \qw &  \ctrl{-1} & \qw  & \qswap \qwx & \qw & \qw   }} \minibox{\hspace{55pt}\Qcircuit @C=.6em @R=.5em {
\lstick{\ket{+}} &\qw &\qw  & \ctrl{1} & \qw & \qw   \\
&\qw  & \qw &  \targ & \qw  & \qw    
}} \minibox{ \hspace{10pt} \large = \hspace{20pt} }
\minibox{\Qcircuit @C=.6em @R=.5em {
\lstick{\ket{+}} &\qw &\qw  & \ctrl{1} & \qw & \qswap & \qw & \qw     \\
&\qw  & \qw &  \targ & \qw  & \qswap \qwx & \qw & \qw   }}
\end{center}
\begin{flushright}
\minibox{\hspace{30pt} \large (S3circ)}
\end{flushright}

\minibox{\Qcircuit @C=.6em @R=.5em {
 &\qw &\qw  & \targ & \qw & \qw & \measureD{Z} & \measure{ get: 0}  \\
&\qw  & \qw &  \ctrl{-1} & \qw  & \qw & \qw   
}} \minibox{ \hspace{7pt} \large =  \hspace{3pt}}
\minibox{\Qcircuit @C=.6em @R=.5em {
 &\qw  & \qswap  & \qw & \targ & \qw & \measureD{Z} & \measure{ get: 0}  \\
&\qw  &  \qswap \qwx & \qw   & \ctrl{-1}  & \qw & \qw }} \minibox{ \hspace{15pt}\Qcircuit @C=.6em @R=.5em {
 &\qw  & \ctrl{1} & \qw & \qw &  \measureD{X} & \measure{ get: +}    \\
&\qw  &  \targ & \qw  & \qw   & \qw   
}} \minibox{  \hspace{7pt} \large =  \hspace{3pt} }\minibox{\Qcircuit @C=.6em @R=.5em {
 &\qw  & \qswap & \qw & \ctrl{1} & \qw &  \measureD{X} & \measure{ get: +}    \\
&\qw  & \qswap \qwx  & \qw  & \targ & \qw & \qw }} \\[0.8in]

\begin{center}
\minibox{\Qcircuit @C=.6em @R=.5em {
\lstick{\ket{0}} &\qw &\qw & \ctrl{1} & \qw & \qw   \\
 &\qw  & \qw &  \targ & \qw    &  \measureD{X} & \measure{ get: +}   
}} \minibox{ \hspace{10pt} \large = \hspace{20pt}}\minibox{\Qcircuit @C=.6em @R=.5em {
 &\qw &\qw & \ctrl{1} & \qw &  \measureD{X} & \measure{ get: +}   \\
\lstick{\ket{0}} &\qw  & \qw &  \targ & \qw  & \qw    
}}  \minibox{ \hspace{10pt} \large = \hspace{20pt}} \minibox{\Qcircuit @C=.6em @R=.5em {
\lstick{\ket{0}}&\qw &\qw  & \targ & \qw  & \qw & \qw    \\
&\qw  & \qw &  \ctrl{-1} & \qw  & \qw & \measureD{X} & \measure{ get: +}  }} \\[0.2in]

\minibox{ \hspace{12pt} \large = \hspace{12pt}} \minibox{\Qcircuit @C=.6em @R=.5em {
&\qw &\qw  & \targ & \qw  & \measureD{X} & \measure{ get: +} \\
\lstick{\ket{0}} &\qw  & \qw &  \ctrl{-1} & \qw  & \qw   }} \minibox{ \hspace{10pt} \large = \hspace{10pt}}\minibox{\Qcircuit @C=.6em @R=.5em {
 &\qw &\qw & \qw & \qw &\qw & \qw &\qw & \qw &\qw & \qw &\qw }}
\end{center} \begin{flushright}
\minibox{\hspace{30pt} \large (S4circ)}
\end{flushright}

\begin{center}
\minibox{\Qcircuit @C=.6em @R=.5em {
\lstick{\ket{+}}&\qw &\qw  & \ctrl{1} & \qw  & \qw & \qw   \\
&\qw  & \qw &  \targ & \qw  & \qw & \measureD{Z} & \measure{ get: 0}  }} \minibox{ \hspace{10pt} \large = \hspace{20pt}} \minibox{\Qcircuit @C=.6em @R=.5em {
&\qw &\qw  & \ctrl{1} & \qw  & \qw & \measureD{Z} & \measure{ get: 0}   \\
\lstick{\ket{+}} &\qw  & \qw &  \targ & \qw  & \qw  & \qw   }}  \minibox{ \hspace{10pt} \large = \hspace{20pt}} \minibox{\Qcircuit @C=.6em @R=.5em {
\lstick{\ket{+}} &\qw &\qw & \targ & \qw & \qw      \\
 &\qw  & \qw &  \ctrl{-1}& \qw  & \measureD{Z} & \measure{ get: 0}   }} \\[0.2in]

 \minibox{ \hspace{12pt} \large = \hspace{12pt}} \minibox{\Qcircuit @C=.6em @R=.5em {
 &\qw &\qw & \targ & \qw &  \measureD{Z} & \measure{ get: 0}   \\
\lstick{\ket{+}} &\qw  & \qw &  \ctrl{-1}& \qw  & \qw    
}}  \minibox{ \hspace{10pt} \large = \hspace{10pt}}  \minibox{\Qcircuit @C=.6em @R=.5em {
 &\qw &\qw & \qw & \qw &\qw & \qw &\qw & \qw &\qw & \qw &\qw }}  \\[0.8in]
\end{center}

\minibox{\Qcircuit @C=.6em @R=.5em {
\lstick{\ket{0}}&\qw &\qw  & \targ &\qw  & \targ & \qw  & \measureD{Z} & \measure{ get: 0}  \\
&\qw  & \qw &  \ctrl{-1} & \qw &  \ctrl{-1} & \qw  & \qw   }} \minibox{ \hspace{20pt} \large = \hspace{30pt}}\minibox{\Qcircuit @C=.6em @R=.5em {
\lstick{\ket{+}}&\qw &\qw  & \ctrl{1} & \qw  & \ctrl{1} & \qw  & \measureD{X} & \measure{ get: +}  \\
&\qw  & \qw &  \targ & \qw &  \targ & \qw  & \qw   }} \minibox{ \hspace{20pt} \large = \hspace{20pt}} \minibox{\Qcircuit @C=.6em @R=.5em {
 &\qw &\qw & \qw & \qw &\qw & \qw &\qw & \qw &\qw & \qw &\qw }} \begin{flushright}
\minibox{\hspace{30pt} \large (S5circ)} \\[0.5in]
\end{flushright}

\begin{center}
\minibox{\Qcircuit @C=.6em @R=.5em {&\qw &\qw & \gate{R_Z(\alpha)} & \qw & \gate{R_Z(\beta)} & \qw &\qw  }} \minibox{ \hspace{20pt} \large = \hspace{20pt}} \minibox{\Qcircuit @C=.6em @R=.5em {&\qw &\qw & \gate{R_Z(\alpha+ \beta)} & \qw &\qw }}
\end{center}
 \begin{flushright}
\minibox{\hspace{30pt} \large (S6circ)} 
\end{flushright}

\begin{center}
\minibox{\Qcircuit @C=.6em @R=.5em {&\qw &\qw & \gate{R_X(\alpha)} & \qw & \gate{R_X(\beta)} & \qw &\qw  }} \minibox{ \hspace{20pt} \large = \hspace{20pt}} \minibox{\Qcircuit @C=.6em @R=.5em {&\qw &\qw & \gate{R_X(\alpha+ \beta)} & \qw &\qw }} \\[0.7in]
\end{center} 

\begin{center}
\minibox{\Qcircuit @C=.6em @R=.5em {
\lstick{\ket{0}} &\qw &\qw & \ctrl{1} & \qw & \qw   \\
&\qw  & \qw &  \targ & \qw  & \qw    
}} \minibox{ \hspace{10pt} \large = \hspace{20pt} }
\minibox{\Qcircuit @C=.6em @R=.5em {
\lstick{\ket{0}} &\qw &\qw   & \qw  & \qw & \qw & \qw   \\
&\qw  & \qw  & \qw  & \qw & \qw & \qw  }} \minibox{\hspace{55pt}\Qcircuit @C=.6em @R=.5em {
 &\qw &\qw  & \ctrl{1} & \qw & \qw   \\
\lstick{\ket{+}} &\qw  & \qw &  \targ & \qw  & \qw    
}} \minibox{ \hspace{10pt} \large = \hspace{20pt} }
\minibox{\Qcircuit @C=.6em @R=.5em {
 &\qw &\qw   & \qw & \qw & \qw & \qw     \\
\lstick{\ket{+}} &\qw  & \qw &  \qw   & \qw & \qw & \qw   }}
\end{center}
\begin{flushright}
\minibox{\hspace{30pt} \large (B1circ)}
\end{flushright}

\minibox{\Qcircuit @C=.6em @R=.5em {
&\qw &\qw & \ctrl{1} & \qw & \qw &  \measureD{Z} & \measure{ get: 0}  \\
&\qw  & \qw &  \targ & \qw  & \qw & \qw    
}} \minibox{ \hspace{5pt} \large = \hspace{5pt} }
\minibox{\Qcircuit @C=.6em @R=.5em {
 &\qw &\qw   & \qw  & \qw & \qw & \measureD{Z} & \measure{ get: 0}    \\
&\qw  & \qw  & \qw  & \qw & \qw & \qw   }} \minibox{\hspace{15pt}\Qcircuit @C=.6em @R=.5em {
 &\qw &\qw  & \ctrl{1} & \qw & \qw & \qw    \\
 &\qw  & \qw &  \targ & \qw  & \qw &   \measureD{X} & \measure{ get: +}     
}} \minibox{ \hspace{5pt} \large = \hspace{5pt} } \minibox{\Qcircuit @C=.6em @R=.5em {
 &\qw &\qw   & \qw & \qw & \qw & \qw     \\
&\qw  & \qw &  \qw   & \qw & \qw &   \measureD{X} & \measure{ get: +}  }}\\[0.5in]

\begin{center}
\minibox{\Qcircuit @C=.6em @R=.5em {
 &\qw &\qw & \ctrl{1} & \qw & \targ & \qw & \qw   \\
&\qw  & \qw &  \targ & \qw &\ctrl{-1}  & \qw  & \qw   
}} \minibox{ \hspace{10pt} \large = \hspace{10pt} }
\minibox{\Qcircuit @C=.6em @R=.5em {
&\qw &\qw   & \targ  & \qw & \qswap & \qw & \qw  \\
&\qw  & \qw  & \ctrl{-1} & \qw & \qswap \qwx & \qw & \qw }} \minibox{ \hspace{10pt} \large = \hspace{10pt} }
\minibox{\Qcircuit @C=.6em @R=.5em {
&\qw &\qw & \qswap   & \qw & \ctrl{1}    & \qw & \qw  \\
&\qw  & \qw & \qswap \qwx &   \qw & \targ   & \qw & \qw }}
\end{center}
\begin{flushright}
\minibox{\hspace{30pt} \large (B2circ) \\[0.5in]}
\end{flushright}

\begin{center}
\minibox{\Qcircuit @C=.6em @R=.5em {
\lstick{\ket{1}} &\qw &\qw & \ctrl{1} & \qw & \qw   \\
&\qw  & \qw &  \targ & \qw  & \qw    
}} \minibox{ \hspace{10pt} \large = \hspace{20pt} }
\minibox{\Qcircuit @C=.6em @R=.5em {
\lstick{\ket{1}} &\qw &\qw   & \qw  & \qw & \qw & \qw   \\
&\qw  & \qw  & \gate{X}  & \qw & \qw & \qw  }} \minibox{\hspace{55pt}\Qcircuit @C=.6em @R=.5em {
 &\qw &\qw  & \ctrl{1} & \qw & \qw   \\
\lstick{\ket{-}} &\qw  & \qw &  \targ & \qw  & \qw    
}} \minibox{ \hspace{10pt} \large = \hspace{20pt} }
\minibox{\Qcircuit @C=.6em @R=.5em {
 &\qw &\qw   & \gate{Z} & \qw & \qw & \qw     \\
\lstick{\ket{-}} &\qw  & \qw &  \qw   & \qw & \qw & \qw   }}
\end{center}
\begin{flushright}
\minibox{\hspace{30pt} \large (K1circ)}
\end{flushright}

\minibox{\Qcircuit @C=.6em @R=.5em {
&\qw &\qw & \ctrl{1} & \qw & \qw &  \measureD{Z} & \measure{ get: 1}  \\
&\qw  & \qw &  \targ & \qw  & \qw & \qw    
}} \minibox{  \large =  }\minibox{\Qcircuit @C=.6em @R=.5em {
 &\qw &\qw   & \qw  & \qw & \qw & \measureD{Z} & \measure{ get: 1}    \\
&\qw  & \qw  & \gate{X}  & \qw & \qw & \qw   }} \minibox{\hspace{15pt}\Qcircuit @C=.6em @R=.5em {
 &\qw &\qw  & \ctrl{1} & \qw & \qw & \qw    \\
 &\qw  & \qw &  \targ & \qw  & \qw &   \measureD{X} & \measure{ get: -}     
}} \minibox{  \large =  } \minibox{\Qcircuit @C=.6em @R=.5em {
 &\qw &\qw   & \gate{Z} & \qw & \qw & \qw     \\
&\qw  & \qw &  \qw   & \qw & \qw &   \measureD{X} & \measure{ get: -}  }}\\[0.9in]

\minibox{\Qcircuit @C=.6em @R=.5em {&\qw &\qw & \gate{Z} & \qw & \gate{R_X(\alpha)} & \qw &\qw  }} \minibox{ \hspace{10pt} \large = \hspace{10pt}} \minibox{\Qcircuit @C=.6em @R=.5em {&\qw &\qw & \gate{R_X(-\alpha)} & \qw & \gate{Z} &\qw &\qw }} \minibox{ \hspace{20pt} \Qcircuit @C=.6em @R=.5em {&\qw &\qw & \gate{X} & \qw & \gate{R_Z(\alpha)} & \qw &\qw  }} \minibox{ \hspace{10pt} \large = \hspace{10pt}} \minibox{\Qcircuit @C=.6em @R=.5em {&\qw &\qw & \gate{R_Z(-\alpha)} & \qw & \gate{X} &\qw &\qw }}  \begin{flushright}\minibox{\hspace{30pt} \large (K2circ)\\[0.4in]} 
\end{flushright}

\begin{center}
\minibox{\Qcircuit @C=1em @R=.5em {
&\qw & \ctrl{1} &\qw &  \qw & \qw & \qw & \qw  & \qw & \qw & \qw & \qw & \qw & \qw & \qw & \qw & \qw & \qw & \qw &   \measureD{X} & \measure{ get: +}   \\
 &\qw  & \targ &\qw & \ctrl{1}  & \qw & \qw & \qw  & \qw & \qw & \qw & \qw & \qw & \qw & \qw & \qw & \qw & \qw & \qw  &   \measureD{X} & \measure{ get: +}  \\
&\qw   & \qw  & \qw  & \targ & \qw & \ctrl{1} & \qw  & \qw & \qw & \qw & \qw & \qw & \qw & \qw & \qw & \qw & \qw & \qw  &  \measureD{X} & \measure{ get: +}  \\
&  & \push{...}  &  & & & \push{...} &  & \push{...} &   & &   & &   & &   & &   & \push{...} & &  \\
&\qw  & \qw &\qw  & \qw &\qw  & \qw & \qw & \targo{-1} & \qw & \gate{R_X(\alpha)}&  \qw & \targo{1} & \qw & \qw & \qw & \qw & \qw & \qw & \qw  \\
&  & \push{...}  &  & & &  &  &  &   &  & &\push{...} & & \push{...} & & & & \push{...}& & \\
\lstick{\ket{+}} &\qw & \qw & \qw & \qw & \qw & \qw & \qw & \qw & \qw & \qw & \qw & \qw & \qw   & \ctrl{-1} & \qw & \targ & \qw & \qw & \qw \\
\lstick{\ket{+}} &\qw & \qw & \qw & \qw & \qw & \qw & \qw & \qw & \qw & \qw & \qw & \qw & \qw   & \qw & \qw & \ctrl{-1} & \qw & \targ & \qw   \\
\lstick{\ket{+}} &\qw & \qw  & \qw & \qw & \qw & \qw & \qw & \qw & \qw & \qw & \qw & \qw & \qw   & \qw & \qw & \qw & \qw & \ctrl{-1}& \qw    
}} \\[0.1in] \minibox{ \hspace{10pt} \large = \hspace{10pt}} \\
\minibox{\Qcircuit @C=1em @R=.5em {
&\qw & \gate{H} &\qw & \targ &\qw &  \qw & \qw & \qw & \qw  & \qw & \qw & \qw & \qw & \qw & \qw & \qw & \qw & \qw & \qw & \qw & \qw & \qw &   \measureD{Z} & \measure{ get: 0}   \\
 &\qw  & \gate{H} & \qw & \ctrl{-1} &\qw & \targ  & \qw & \qw & \qw  & \qw & \qw & \qw & \qw & \qw & \qw & \qw & \qw & \qw & \qw & \qw & \qw & \qw  &   \measureD{Z} & \measure{ get: 0}  \\
&\qw & \gate{H} & \qw  & \qw  & \qw  & \ctrl{-1} & \qw & \targo{1} & \qw  & \qw & \qw & \qw & \qw & \qw & \qw & \qw & \qw & \qw & \qw & \qw & \qw & \qw &  \measureD{Z} & \measure{ get: 0}  \\
&  & \push{...}  & &   &  & & & \push{...} &  & \push{...} &   & &   & &   & &   & & & &  & \push{...} & &  \\
&\qw  & \qw & \qw & \qw &\qw  & \qw &\qw  & \qw & \qw & \ctrl{-1} & \qw & \gate{R_Z(\alpha)}&  \qw & \ctrl{1} & \qw & \qw & \qw & \qw & \qw & \qw & \qw & \qw & \qw  \\
&  & \push{...}  &  & & & &  &  &  &  &   &  & &\push{...} & & \push{...} & & & & & & \push{...}& & \\
\lstick{\ket{0}} &\qw & \qw & \qw & \qw & \qw & \qw & \qw & \qw & \qw & \qw & \qw & \qw & \qw & \qw & \qw   & \targo{-1} & \qw & \ctrl{1} & \qw & \qw & \qw & \gate{H} &\qw  \\
\lstick{\ket{0}} &\qw & \qw & \qw & \qw & \qw & \qw & \qw & \qw & \qw & \qw & \qw & \qw & \qw & \qw & \qw   & \qw & \qw & \targ & \qw & \ctrl{1} & \qw & \gate{H} &\qw   \\
\lstick{\ket{0}} &\qw  & \qw & \qw& \qw  & \qw & \qw & \qw & \qw & \qw & \qw & \qw & \qw & \qw & \qw & \qw   & \qw & \qw & \qw & \qw & \targ & \qw & \gate{H} &\qw    
}}
\end{center} \begin{flushright}\minibox{\hspace{30pt} \large (Ccirc)\\[0.5in]} 
\end{flushright}
Note that this rule also holds if both sides of the (Ccirc) equation above only contain the top/bottom half of the quantum circuit (corresponding to the (C) rule with no inputs/outputs respectively). \\[0.6in]

\begin{center}
\minibox{\Qcircuit @C=.6em @R=.5em {&\qw &\qw & \qw & \gate{H} & \qw & \qw &\qw  }} \minibox{ \hspace{20pt} \large = \hspace{20pt}} \minibox{\Qcircuit @C=.6em @R=.5em {&\qw &\qw & \gate{R_Z(\frac{\pi}{2})} & \qw & \gate{R_X(\frac{\pi}{2})}&\qw & \gate{R_Z(\frac{\pi}{2})} &\qw &\qw }} 
\end{center}
\begin{flushright}\minibox{\hspace{30pt} \large (Hcirc) \\[0.4in]} 
\end{flushright}

\begin{figure}[ht!]
Let us associate a number to each input and output of a quantum circuit Q. If we can obtain a valid quantum circuit Q', whose inputs and outputs are numbered in the same way as Q, by replacing a finite number of times the following quantum circuit fragments: \\[0.12in]

\minibox{\Qcircuit @C=1em @R=.5em {
 \lstick{\ket{+}} &\qw &\qw  & \ctrl{1} & \qw & \qw   \\
 & &  &  \push{\cdots} &  &   
}} \minibox{\hspace{7pt}\large ; \hspace{20pt}}\minibox{\Qcircuit @C=1em @R=.5em {
  & &  &  \push{\cdots} &  &   \\
 \lstick{\ket{0}} &\qw &\qw  & \targo{-1} & \qw & \qw  
}} \minibox{\hspace{7pt}\large ; \hspace{10pt}} \minibox{\Qcircuit @C=1em @R=.5em {
&\qw &\qw  & \ctrl{1} & \qw & \qw  & \measureD{X} & \measure{ get: +}    \\
 & &  &  \push{\cdots} &  &   & &
}} \minibox{\hspace{7pt}\large ; \hspace{10pt}} \minibox{\Qcircuit @C=1em @R=.5em {
 &  &  &  \push{\cdots}  &  &  & \\
 &\qw &\qw  & \targo{-1} & \qw & \qw & \measureD{Z} & \measure{ get: 0}  
}} \\[0.12in]

by wires with the same number as the corresponding input or output (regardless of topological structure),\\ then the circuits Q and Q' are equivalent.  \hspace{0.3in} (Scirc)

For example, the following circuit equation follows from the application of the (Scirc) rule: \\[0.2in]

\begin{center}
\minibox{\Qcircuit @C=.6em @R=.5em {
  & \lstick{1}  &\qw &\qw & \qw& \qw & \targ & \qw &  \measureD{Z} & \measure{ get: 0}   \\
 & \lstick{2}   &\qw & \targ & \qw &  \qw & \ctrl{-1} &  \qw & \qw  & \rstick{3}   \\
\lstick{\ket{+}}  &\qw & \qw &  \ctrl{-1} &  \qw & \qw  & \qw & \qw  & \qw  &\rstick{4}     
}}  \minibox{ \hspace{15pt} \large = \hspace{15pt}} \minibox{\Qcircuit @C=.6em @R=.5em {
& \lstick{1} &\qw &\qw & \qw & \ctrl{1} & \qw  & \qw & \qw  & \rstick{3}  \\
 & \lstick{2} &\qw  & \qw & \qw &  \targ & \qw  & \qw & \qw & \rstick{4} }} \\[0.8in]
\end{center}
 
\caption{Sound and complete set of circuit equations for stabilizer quantum mechanics.}
\label{stabcirc}
\end{figure}

\end{widetext}

Therefore, we have found a \textbf{complete set of quantum circuit equations for stabilizer quantum mechanics}. Any circuit equation which can be shown to be true using stabilizer theory---in the sense that both quantum circuits in the equation correspond to equivalent processes in stabilizer quantum mechanics---can be derived from this set. This provides a novel insight into the logical foundation of the stabilizer formalism. \\

\section{Derivation of an equation between stabilizer quantum circuits from the complete set}

The proof of the result relies heavily upon categorical quantum mechanics. It would have been difficult to find this set of circuits without the flexibility of the ZX network and the theorem may have been difficult to prove without appealing to category theory.

The theorem itself, however, is purely a result about quantum circuits and stabilizer quantum mechanics, which can readily be understood without any knowledge of category theory or formal logic. 

In order to make this clear and provide an illustration of the general result, we now give an example of using the complete set of circuit equations to formally derive a well known equation between stabilizer quantum circuits.  

The first quantum circuit of the equation below corresponds to the standard quantum teleportation protocol \cite{Be93}, where a Bell state $\ket{00}+\ket{11}$ is prepared on the second and third qubits and the Bell basis is measured on the first two qubits (the result corresponding to $\ket{00}+\ket{11}$ is post-selected). We use the complete set of circuit equations from Figure \ref{stabcirc} to show that this is the same quantum process as taking the first qubit to the third qubit:

\[
\Qcircuit @C=1.0em @R=.7em {
& \qw & \qw & \qw & \qw & \qw & \qw & \ctrl{1} & \qw & \measureD{X} & \measure{ get: +}\\
& & & \lstick{\ket{+}} & \qw & \ctrl{1} & \qw & \targ & \qw & \qw & \qw  & \qw & \measureD{Z} & \measure{ get: 0} \\
 \lstick{\ket{0}} & \qw  & \qw & \qw & \qw & \targ & \qw & \qw & \qw \gategroup{1}{2}{3}{11}{1.0em}{--} & \qw  & \qw & \qw & \qw & \qw  
}
\] 
\begin{center}
\large = \\
\small (S2circ)
\end{center} 
 \[
\Qcircuit @C=1.0em @R=.7em {
& \qw & \qw & \qw & \qw & \qw & \ctrl{1} & \qw  & \qw & \measureD{X} & \measure{ get: +}\\
\lstick{\ket{0}} & \qw  & \qw & \qw & \qw & \qw & \targ & \qw & \targ & \qw & \qw  & \qw & \measureD{Z} & \measure{ get: 0} \\
& & & \lstick{\ket{+}} & \qw & \qw  & \qw & \qw & \ctrl{-1}  \gategroup{1}{2}{3}{11}{1.0em}{--} & \qw  & \qw & \qw & \qw & \qw  
}
\]

\begin{center}
\large = \\
\small (Ccirc)
\end{center}
 \[
\Qcircuit @C=0.8em @R=.6em {
& \qw & \qw & \qw  & \gate{H} & \qw & \targ & \qw  & \qw & \qw  & \measureD{Z} & \measure{ get: 0}\\
\lstick{\ket{0}} & \qw  & \qw & \qw   & \gate{H} & \qw & \ctrl{-1} & \qw & \ctrl{1} & \qw &  \gate{H} & \qw  & \qw & \measureD{Z} & \measure{ get: 0} \\
& & & \lstick{\ket{0}} & \qw  &  \qw & \qw & \qw & \targ &  \qw & \gate{H}  & \qw \gategroup{1}{2}{3}{12}{1.1em}{--}  & \qw & \qw  
}
\]

\begin{center}
\large = \\
\small (Ccirc)
\end{center}
 \[
\Qcircuit @C=0.85em @R=.58em {
& \qw & \gate{H} & \qw  & \qw & \qw & \qw & \targ  & \measureD{Z} & \measure{ get: 0}\\
 & & & & &\lstick{\ket{+}} & \qw  & \ctrl{-1} & \qw & \qw & \ctrl{1} & \qw  & \measureD{X} & \measure{ get: +} \\
& & & & & \lstick{\ket{0}} & \qw & \qw  & \qw & \qw & \targ &  \qw & \gate{H}  & \qw  \gategroup{1}{4}{2}{10}{1.0em}{--}  
}
\]

\begin{center}
\large = \\
\small (S4circ)
\end{center}
 \[
\Qcircuit @C=1.0em @R=.7em {
 & \qw & \gate{H} & \qw & \qw & \ctrl{1} & \qw  & \measureD{X} & \measure{ get: +} \\
 & \lstick{\ket{0}} & \qw  & \qw & \qw & \targ &  \qw & \gate{H} & \qw \gategroup{1}{4}{1}{5}{1.0em}{--}  
}
\]

\begin{center}
\large = \\
\small (S4circ)
\end{center}
 \[
\Qcircuit @C=1.0em @R=.7em {
 & \qw & \gate{H} & \qw & \gate{H} & \qw \gategroup{1}{3}{1}{3}{1.0em}{--}  
}
\]

\begin{center}
\large = \\
\small (Hcirc)
\end{center}
\scalebox{0.87}{\minibox{
\Qcircuit @C=0.5em @R=.35em {
 & \qw & \gate{R_Z(\frac{\pi}{2})} & \qw & \gate{R_X(\frac{\pi}{2})} & \qw & \gate{R_Z(\frac{\pi}{2})} 
 & \qw & \gate{R_Z(\frac{\pi}{2})} & \qw  & \gate{R_X(\frac{\pi}{2})} & \qw & \gate{R_Z(\frac{\pi}{2})} & \qw \gategroup{1}{3}{1}{7}{0.6em}{--} 
}}}

\begin{center}
\large = \\
\small (S6circ)
\end{center}

\scalebox{0.9}{\minibox{\Qcircuit @C=0.5em @R=.35em {
 & \qw & \gate{R_Z(\frac{\pi}{2})} & \qw & \gate{R_X(\frac{\pi}{2})} & \qw & \gate{R_Z(\pi)} & \qw  & \gate{R_X(\frac{\pi}{2})} & \qw & \gate{R_Z(\frac{\pi}{2})} & \qw \gategroup{1}{7}{1}{9}{0.6em}{--} 
}}}

\begin{center}
\large = \\
\small (K2circ)
\end{center}
 
\scalebox{0.9}{\minibox{\Qcircuit @C=0.5em @R=.35em {
 & \qw & \gate{R_Z(\frac{\pi}{2})} & \qw & \gate{R_X(\frac{\pi}{2})} & \qw  & \gate{R_X(\frac{-\pi}{2})} & \qw & \gate{R_Z(\pi)} & \qw & \gate{R_Z(\frac{\pi}{2})} & \qw \gategroup{1}{7}{1}{9}{0.6em}{--} 
}}}

\begin{center}
\large = \\
\small (S6circ),(K2circ) 
\end{center} \[
\Qcircuit @C=1.0em @R=.7em {
 & \qw & \qw & \qw & \qw & \qw & \qw & \qw & \qw & \qw & \qw & \qw & \qw & \qw & \qw & \qw & \qw & \qw & \qw & \qw & \qw & \qw
 }
\] \\

This is a proof of the validity of quantum teleportation from a set of axioms for quantum stabilizer theory. The dotted boxes indicate a circuit substitution using a circuit equation from Figure \ref{stabcirc}. Any equivalence between two quantum circuits corresponding to the same stabilizer process can be formally shown from the complete set of circuit equations by using this reasoning by substitution.

\section{Reasoning with the ZX network is much easier than with the quantum circuit calculus}

A quick comparison of the ZX network axioms from Figure \ref{ZXRules} with the set of quantum circuit axioms from Figure \ref{stabcirc} makes it clear that demonstrating the equivalence of quantum processes with the quantum circuit calculus will be far more cumbersome than using the ZX network. For instance, in the previous section, the circuit calculus takes more than 10 steps to prove the validity of the post-selected teleportation protocol, whereas the ZX network can verify validity in a single step.

Now, let us briefly present another example of a derivation which is less trivial using the ZX network. This demonstrates how the flexibility of the spider law allows the ZX network to show validity of a quantum circuit equation far more intuitively and efficiently than the quantum circuit calculus. Both the ZX network and the quantum circuit calculus can prove that the following measurement based quantum computing program computes a CNOT gate: 

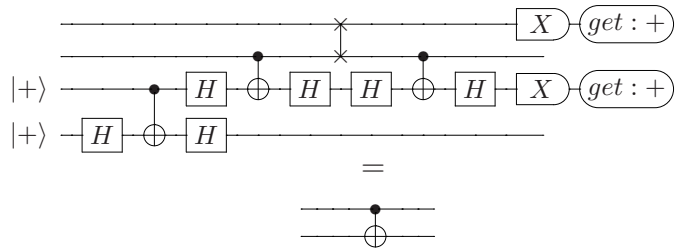
\begin{center}
\minibox{\Qcircuit @C=.4em @R=.4em {
 &\qw & \qw &\qw & \qw  & \qw & \qw & \qw &  \qw & \qw & \qw & \qswap  & \qw  & \qw & \qw & \qw & \qw & \qw  & \measureD{X} & \measure{ get: +}  \\
& \qw & \qw & \qw & \qw  & \qw  & \qw & \qw &  \ctrl{1} & \qw &  \qw &  \qswap \qwx   & \qw & \qw  & \ctrl{1} & \qw & \qw & \qw  & \qw  \\
\lstick{\ket{+}} &\qw & \qw &\qw  & \ctrl{1} & \qw & \gate{H} & \qw & \targ & \qw &  \gate{H} & \qw  & \gate{H} & \qw& \targ & \qw & \gate{H} & \qw & \measureD{X} & \measure{ get: +} \\
\lstick{\ket{+}} &\qw & \gate{H} &\qw  & \targ &  \qw & \gate{H} &  \qw & \qw  & \qw & \qw & \qw & \qw & \qw & \qw & \qw & \qw & \qw & \qw   }}  
\minibox{ \hspace{30pt} \large = \hspace{25pt}} \\
\minibox{\Qcircuit @C=.6em @R=.5em {
&\qw &\qw &\qw & \ctrl{1} & \qw   & \qw  & \qw  \\
&\qw  & \qw  &\qw & \targ & \qw  & \qw  & \qw   }}
\end{center} 

This only requires a straightforward repeated application of the (S) law and 2 applications of the (C) law using the ZX network \cite{Du08}. The circuit calculus, however, requires applications of the (Hcirc), (S6circ), (K2circ), (Ccirc), (S2circ), (S3circ) and (Scirc) rules to demonstrate the validity of the previous equation. Therefore, using the circuit calculus to check correctness not only requires a larger total number of axioms to be used but also uses more distinct axioms, whose application is far less intuitive than in the ZX network case.

The examples presented above are circuit equations whose validity can be shown in a small number of steps. For larger circuit equations, we expect the use of the circuit calculus to be unviable. The skeptical reader is challenged to verify the correctness of the 7 qubit Steane code \cite{Luc13} using the circuit calculus instead of the ZX network.  

We conclude this section by stressing once again that the elements of the ZX network have no explicit physical meaning. Indeed, the network elements are not restricted to the circuit structure of quantum processes. This mathematical flexibility is at the core of the calculational power of the network calculus relative to the circuit calculus. For example, a primitive circuit element like the CNOT gate is broken down into two abstract elements in the ZX network, corresponding to red and green nodes. These elements obey algebraic rules, some of which have no evident physical interpretation, but which appear to play a fundamental logical role. In contrast, every rule in the circuit calculus has an explicit physical interpretation.

\section{Conclusion}

Studying quantum theory from a logical, computer science perspective has provided an insight into the foundations of stabilizer quantum mechanics. The axiomatic approach presented here provides a representation of the systems and processes of an operational physical theory, together with all the equational laws they obey.

Describing physical processes directly using a logical language may dispense with the need of a more elaborate mathematical description which would require a more refined language and further axioms. Some of this extra structure may be unnecessary and undesirable to fully model an operational physical theory and may even lead to several redundant mathematical descriptions of a single physical theory, like the use of either Hilbert spaces or the ZX network to describe stabilizer quantum mechanics. 

Furthermore, such a formalization of the foundations of physics allows one to rigorously ask certain questions about consistency, soundness and completeness of physical theories. Is it possible to find a consistent, sound and complete set of quantum circuit equations which can prove the validity of any true quantum circuit equation? Are there fundamental incompleteness theorems for the foundations of physics? \\

In any case, the study of the logical foundation of physical theories is an essential method of testing their validity, especially in realms of nature in which experiments are very difficult or impossible to perform. Logic seems to be the most suited tool to rigorously study the foundations of mathematical theories of nature from a human perspective.

\section*{Acknowledgments}

Terry Rudolph's insightful remarks provided some of the initial motivation for this work. We would also like to thank Miriam Backens and Aleks Kissinger for helpful discussions. We acknowledge financial support from the EPSRC.

\bibliography{Mresbib}

\newpage

\appendix

\section{}

\begin{figure}[H]
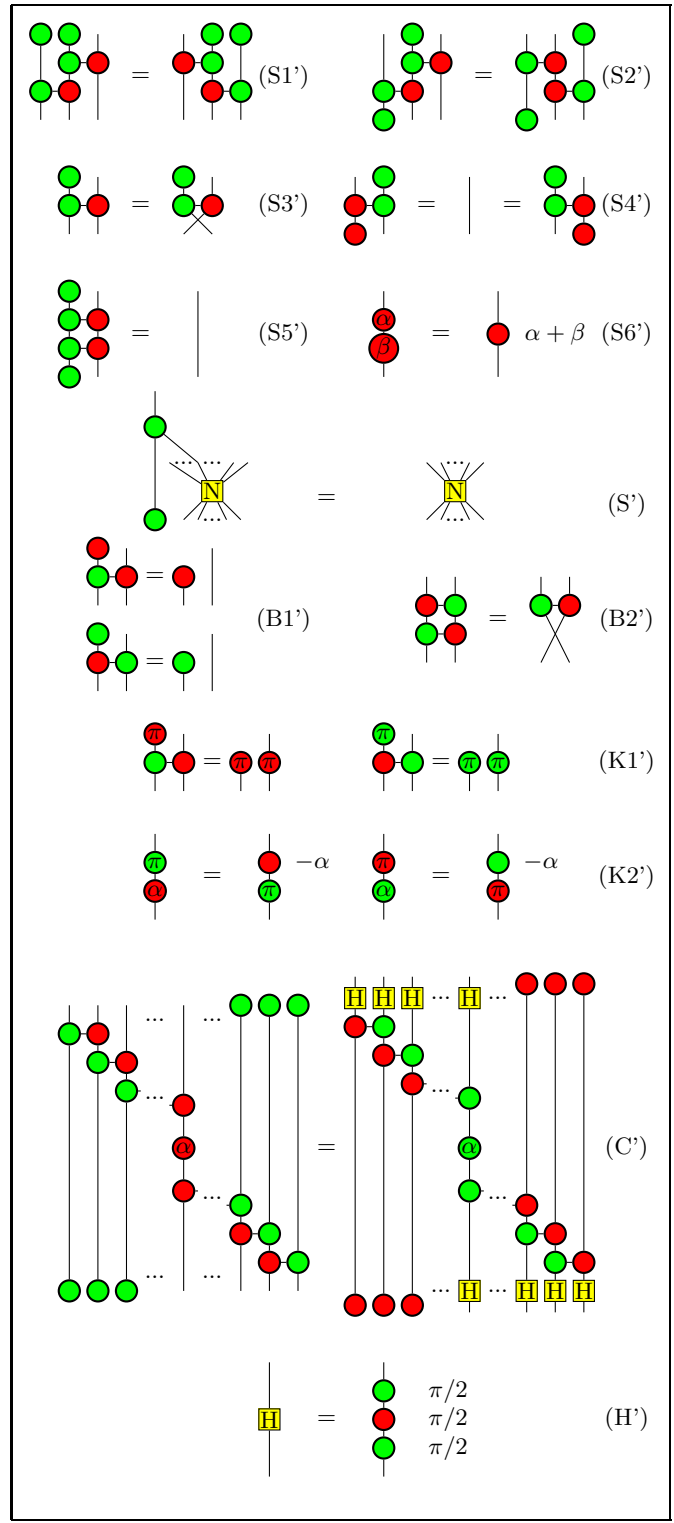


\frame{
}
\caption{Alternative ZX axioms in a form resembling quantum circuit equations.}
\label{Circax}	
\end{figure}

We will now prove that the set of ZX network equations given in Figure \ref{Circax}, which are in a form that can be directly related to quantum circuits using Figure \ref{Circ}, are equivalent to the axioms of the ZX network. Note that normalization is not relevant for the proof of completeness so we ignore scalar factors. \\

Note first of all that the rule (T) of the ZX network states that after enumerating the inputs and outputs of a diagram, any topological deformation of the internal structure will give an equal diagram. The (T) rule can be used as part of the new set of ZX axioms in the form resembling circuit equations. The topological rigidity of quantum circuits, however, means that the complete set of quantum circuit equations will contain several equations for each ZX network rule, one for each possible choice of assignments of inputs and outputs.  \\ 

\textbf{Lemma A1}: The ZX network rules (S1'), (S2'), (S3'), (S4'), (S5'), (S6') and (S') taken together are equivalent to the (S) rules of the ZX network: \\

\begin{tikzpicture}[scale=0.47]
	\begin{pgfonlayer}{nodelayer}
		\node [style=rn] (0) at (-5, 8) {$\alpha$};
		\node [style=rn] (1) at (-3, 6) {$\beta$};
		\node [style=none] (2) at (-6, 9) {};
		\node [style=none] (3) at (-4, 9) {};
		\node [style=none] (4) at (-4, 5) {};
		\node [style=none] (5) at (-2, 5) {};
		\node [style=none] (6) at (-3, 5) {...};
		\node [style=none] (7) at (-4, 7) {...};
		\node [style=none] (8) at (-5, 9) {...};
		\node [style=none] (9) at (0, 7) {=};
		\node [style=rn] (10) at (3, 7) {};
		\node [style=none] (11) at (2, 9) {};
		\node [style=none] (12) at (4, 9) {};
		\node [style=none] (13) at (2, 5) {};
		\node [style=none] (14) at (4, 5) {};
		\node [style=none] (15) at (-6.5, 5) {};
		\node [style=none] (16) at (-5, 5) {};
		\node [style=none] (17) at (-3, 9) {};
		\node [style=none] (18) at (-1.5, 9) {};
		\node [style=none] (19) at (-2.25, 9) {...};
		\node [style=none] (20) at (-5.75, 5) {...};
		\node [style=none] (21) at (3, 5) {...};
		\node [style=none] (22) at (3, 9) {...};
		\node [style=none] (23) at (9.500001, 7) {(S1)};
		\node [style=none] (24) at (0, -2) {$\Leftrightarrow$};
		\node [style=none] (25) at (-7.5, 0.7499999) {};
		\node [style=none] (26) at (2, -0.4999999) {};
		\node [style=rn] (27) at (2, 1.5) {};
		\node [style=none] (28) at (-5.5, 0.7499999) {};
		\node [style=none] (29) at (5, -0.4999999) {};
		\node [style=none] (30) at (2, 3.5) {};
		\node [style=none] (31) at (5, 3.5) {};
		\node [style=none] (32) at (-3, 0.7499999) {};
		\node [style=none] (33) at (-0.9999999, 0.7499999) {};
		\node [style=rn] (34) at (-6.5, 2) {};
		\node [style=none] (35) at (-4.25, 1.5) {=};
		\node [style=none] (36) at (3.5, 1.5) {=};
		\node [style=none] (37) at (9.500001, 1.5) {(S2)};
		\node [style=none] (38) at (4.7, 7) {$\alpha+\beta$};
	\end{pgfonlayer}
	\begin{pgfonlayer}{edgelayer}
		\draw (3.center) to (0);
		\draw (2.center) to (0);
		\draw [style=simple, bend left=45, looseness=1.25] (0) to (1);
		\draw [style=simple, bend right=45, looseness=1.50] (0) to (1);
		\draw  (1) to (4.center);
		\draw  (1) to (5.center);
		\draw  (11.center) to (10);
		\draw  (10) to (12.center);
		\draw  (10) to (14.center);
		\draw  (10) to (13.center);
		\draw [style=simple, bend right=15, looseness=1.00] (1) to (17.center);
		\draw  (1) to (18.center);
		\draw  (0) to (15.center);
		\draw [style=simple, bend right=15, looseness=1.00] (0) to (16.center);
		\draw [style=simple, bend left=45, looseness=1.00] (25.center) to (34);
		\draw [style=simple, bend left=45, looseness=1.00] (34) to (28.center);
		\draw [style=simple, bend left=60, looseness=2.50] (32.center) to (33.center);
		\draw  (30.center) to (27);
		\draw  (27) to (26.center);
		\draw  (31.center) to (29.center);
	\end{pgfonlayer}
\end{tikzpicture}

\begin{tikzpicture}[scale=0.39]
	\begin{pgfonlayer}{nodelayer}
		\node [style=gn] (0) at (-9, 7) {};
		\node [style=gn] (1) at (-9, 6) {};
		\node [style=gn] (2) at (-10, 7) {};
		\node [style=gn] (3) at (-10, 5) {};
		\node [style=rn] (4) at (-9, 5) {};
		\node [style=rn] (5) at (-8, 6) {};
		\node [style=none] (6) at (-8, 7) {};
		\node [style=none] (7) at (-8, 4) {};
		\node [style=none] (8) at (-9, 4) {};
		\node [style=none] (9) at (-10, 4) {};
		\node [style=none] (10) at (-6.5, 5.5) {=};
		\node [style=rn] (11) at (-5, 6) {};
		\node [style=gn] (12) at (-4, 6) {};
		\node [style=gn] (13) at (-4, 7) {};
		\node [style=gn] (14) at (-3, 7) {};
		\node [style=gn] (15) at (-3, 5) {};
		\node [style=rn] (16) at (-4, 5) {};
		\node [style=none] (17) at (-5, 4) {};
		\node [style=none] (18) at (-4, 4) {};
		\node [style=none] (19) at (-3, 4) {};
		\node [style=none] (20) at (-5, 7) {};
		\node [style=none] (21) at (-1.5, 5.5) {(S1')};
		\node [style=rn] (22) at (4, 6) {};
		\node [style=rn] (23) at (3, 5) {};
		\node [style=gn] (24) at (3, 6) {};
		\node [style=gn] (25) at (2, 5) {};
		\node [style=gn] (26) at (2, 4) {};
		\node [style=gn] (27) at (3, 7) {};
		\node [style=none] (28) at (2, 7) {};
		\node [style=none] (29) at (4, 7) {};
		\node [style=none] (30) at (3, 4) {};
		\node [style=none] (31) at (4, 4) {};
		\node [style=none] (32) at (5.499999, 5.5) {=};
		\node [style=rn] (33) at (7.999999, 5) {};
		\node [style=rn] (34) at (7.999999, 6) {};
		\node [style=gn] (35) at (9.000001, 5) {};
		\node [style=gn] (36) at (6.999999, 6) {};
		\node [style=gn] (37) at (9.000001, 7) {};
		\node [style=gn] (38) at (6.999999, 4) {};
		\node [style=none] (39) at (7.999999, 4) {};
		\node [style=none] (40) at (9.000001, 4) {};
		\node [style=none] (41) at (6.999999, 7) {};
		\node [style=none] (42) at (7.999999, 7) {};
		\node [style=none] (43) at (10.5, 5.5) {(S2')};
		\node [style=none] (44) at (2, -0) {};
		\node [style=none] (45) at (0.9999999, 2) {};
		\node [style=none] (46) at (3.5, 0.9999999) {=};
		\node [style=none] (47) at (5, 2) {};
		\node [style=none] (48) at (5, -0) {};
		\node [style=none] (49) at (6.499999, 0.9999999) {=};
		\node [style=none] (50) at (7.999999, -0) {};
		\node [style=none] (51) at (9.000001, 2) {};
		\node [style=gn] (52) at (2, 2) {};
		\node [style=gn] (53) at (2, 0.9999999) {};
		\node [style=rn] (54) at (0.9999999, 0.9999999) {};
		\node [style=rn] (55) at (0.9999999, -0) {};
		\node [style=gn] (56) at (7.999999, 2) {};
		\node [style=gn] (57) at (7.999999, 0.9999999) {};
		\node [style=rn] (58) at (9.000001, 0.9999999) {};
		\node [style=rn] (59) at (9.000001, -0) {};
		\node [style=none] (60) at (10.5, 0.9999999) {(S4')};
		\node [style=rn] (61) at (-8, 0.9999999) {};
		\node [style=gn] (62) at (-9, 0.9999999) {};
		\node [style=gn] (63) at (-9, 2) {};
		\node [style=none] (64) at (-8, -0) {};
		\node [style=none] (65) at (-9, -0) {};
		\node [style=none] (66) at (-8, 2) {};
		\node [style=none] (67) at (-5, -0) {};
		\node [style=none] (68) at (-6.5, 0.9999999) {=};
		\node [style=none] (69) at (-4, -0) {};
		\node [style=none] (70) at (-4, 2) {};
		\node [style=gn] (71) at (-5, 2) {};
		\node [style=gn] (72) at (-5, 0.9999999) {};
		\node [style=rn] (73) at (-4, 0.9999999) {};
		\node [style=none] (74) at (-1.5, 0.9999999) {(S3')};
		\node [style=none] (75) at (-8, -2) {};
		\node [style=none] (76) at (-8, -5) {};
		\node [style=none] (77) at (-6.5, -3.5) {=};
		\node [style=none] (78) at (-4.5, -2) {};
		\node [style=none] (79) at (-4.5, -5) {};
		\node [style=none] (80) at (-1.5, -3.5) {(S5')};
		\node [style=none] (81) at (2, -2) {};
		\node [style=none] (82) at (2, -5) {};
		\node [style=none] (83) at (5.999999, -2) {};
		\node [style=none] (84) at (5.999999, -5) {};
		\node [style=rn] (85) at (2, -3) {$\alpha$};
		\node [style=rn] (86) at (2, -4) {$\beta$};
		\node [style=rn] (87) at (5.999999, -3.5) {};
		\node [style=none] (88) at (4, -3.5) {=};
		\node [style=none] (89) at (10.5, -3.5) {(S6')};
		\node [style=gn] (90) at (-9, -2) {};
		\node [style=gn] (91) at (-9, -3) {};
		\node [style=gn] (92) at (-9, -4) {};
		\node [style=gn] (93) at (-9, -5) {};
		\node [style=rn] (94) at (-8, -3) {};
		\node [style=rn] (95) at (-8, -4) {};
		\node [style=none] (96) at (7.7, -3.5) {$\alpha+\beta$};
		\node [style=none] (97) at (4.25, -7.5) {...};
		\node [style=none] (98) at (-4.25, -7.5) {...};
		\node [style=none] (99) at (-3.25, -9.500001) {};
		\node [style=none] (100) at (0, -8.499999) {=};
		\node [style=none] (101) at (4.25, -9.500001) {...};
		\node [style=Had] (102) at (-4.25, -8.499999) {N};
		\node [style=none] (103) at (3.750001, -7.5) {};
		\node [style=none] (104) at (-3.750001, -9.500001) {};
		\node [style=none] (105) at (-5.75, -7.5) {};
		\node [style=Had] (106) at (4.25, -8.499999) {N};
		\node [style=none] (107) at (3.25, -9.500001) {};
		\node [style=none] (108) at (-6.25, -5.499999) {};
		\node [style=none] (109) at (-3, -7.5) {};
		\node [style=none] (110) at (-4.75, -9.500001) {};
		\node [style=none] (111) at (3.25, -7.5) {};
		\node [style=none] (112) at (4.75, -7.5) {};
		\node [style=none] (113) at (-3.500001, -7.5) {};
		\node [style=gn] (114) at (-6.25, -9.500001) {};
		\node [style=none] (115) at (5.249999, -7.5) {};
		\node [style=none] (116) at (3.750001, -9.500001) {};
		\node [style=gn] (117) at (-6.25, -6.5) {};
		\node [style=none] (118) at (-5.250001, -7.5) {...};
		\node [style=none] (119) at (4.75, -9.500001) {};
		\node [style=none] (120) at (-4.75, -7.5) {};
		\node [style=none] (121) at (5.249999, -9.500001) {};
		\node [style=none] (122) at (10.5, -8.499999) {(S')};
		\node [style=none] (123) at (-5.250001, -9.500001) {};
		\node [style=none] (124) at (-4.25, -9.500001) {...};
	\end{pgfonlayer}
	\begin{pgfonlayer}{edgelayer}
		\draw  (0) to (1);
		\draw  (6.center) to (5);
		\draw  (5) to (1);
		\draw  (5) to (7.center);
		\draw  (1) to (4);
		\draw  (4) to (8.center);
		\draw  (3) to (9.center);
		\draw  (4) to (3);
		\draw  (3) to (2);
		\draw  (14) to (15);
		\draw  (15) to (19.center);
		\draw  (16) to (18.center);
		\draw  (16) to (15);
		\draw  (12) to (16);
		\draw  (13) to (12);
		\draw  (11) to (12);
		\draw  (20.center) to (11);
		\draw  (11) to (17.center);
		\draw  (29.center) to (22);
		\draw  (22) to (31.center);
		\draw  (23) to (30.center);
		\draw  (24) to (23);
		\draw  (27) to (24);
		\draw  (24) to (22);
		\draw  (25) to (23);
		\draw  (25) to (26);
		\draw  (28.center) to (25);
		\draw  (41.center) to (36);
		\draw  (34) to (42.center);
		\draw  (37) to (35);
		\draw  (35) to (40.center);
		\draw  (33) to (39.center);
		\draw  (33) to (35);
		\draw  (34) to (36);
		\draw  (34) to (33);
		\draw  (36) to (38);
		\draw  (54) to (53);
		\draw  (52) to (53);
		\draw  (54) to (55);
		\draw  (45.center) to (54);
		\draw  (53) to (44.center);
		\draw  (57) to (50.center);
		\draw  (58) to (59);
		\draw  (58) to (57);
		\draw  (51.center) to (58);
		\draw  (57) to (56);
		\draw  (47.center) to (48.center);
		\draw  (63) to (62);
		\draw  (66.center) to (61);
		\draw  (62) to (61);
		\draw  (61) to (64.center);
		\draw  (62) to (65.center);
		\draw  (71) to (72);
		\draw  (70.center) to (73);
		\draw  (72) to (73);
		\draw  (72) to (69.center);
		\draw  (73) to (67.center);
		\draw  (90) to (91);
		\draw  (91) to (92);
		\draw  (92) to (93);
		\draw  (76.center) to (95);
		\draw  (95) to (92);
		\draw  (94) to (95);
		\draw  (94) to (91);
		\draw  (94) to (75.center);
		\draw  (78.center) to (79.center);
		\draw  (81.center) to (85);
		\draw  (85) to (86);
		\draw  (86) to (82.center);
		\draw  (83.center) to (87);
		\draw  (87) to (84.center);
		\draw (102) to (104.center);
		\draw (102) to (113.center);
		\draw (102) to (109.center);
		\draw (102) to (99.center);
		\draw (102) to (110.center);
		\draw (102) to (123.center);
		\draw (102) to (120.center);
		\draw (105.center) to (102);
		\draw (120.center) to (117);
		\draw (117) to (114);
		\draw (108.center) to (117);
		\draw (106) to (119.center);
		\draw (106) to (112.center);
		\draw (106) to (115.center);
		\draw (106) to (121.center);
		\draw (106) to (116.center);
		\draw (106) to (107.center);
		\draw (106) to (111.center);
		\draw (103.center) to (106);
	\end{pgfonlayer}
\end{tikzpicture}
This equivalence assumes that the (T) rule holds and that the (C) rule holds in one direction.

\textbf{Proof:} By theorems 6.11 and 6.12 of \cite{Du08}, we know that (S1) and (S2) are equivalent to: 

\begin{tikzpicture}[scale=0.37]
	\begin{pgfonlayer}{nodelayer}
		\node [style=gn] (0) at (-9, 6) {};
		\node [style=gn] (1) at (-9.5, 5) {};
		\node [style=none] (2) at (-9, 7) {};
		\node [style=none] (3) at (-8, 4) {};
		\node [style=none] (4) at (-9, 4) {};
		\node [style=none] (5) at (-10, 4) {};
		\node [style=none] (6) at (-6.5, 5.5) {=};
		\node [style=gn] (7) at (-4, 6) {};
		\node [style=gn] (8) at (-3.5, 5) {};
		\node [style=none] (9) at (-5, 4) {};
		\node [style=none] (10) at (-4, 4) {};
		\node [style=none] (11) at (-3, 4) {};
		\node [style=none] (12) at (-4, 7) {};
		\node [style=none] (13) at (-1.5, 5.5) {(S1o')};
		\node [style=rn] (14) at (3.000001, 6) {};
		\node [style=rn] (15) at (2, 5) {};
		\node [style=none] (16) at (1, 7) {};
		\node [style=none] (17) at (3.000001, 7) {};
		\node [style=none] (18) at (2, 4) {};
		\node [style=none] (19) at (4.000001, 4) {};
		\node [style=none] (20) at (5.499999, 5.5) {=};
		\node [style=rn] (21) at (7.999999, 5) {};
		\node [style=rn] (22) at (7.999999, 6) {};
		\node [style=none] (23) at (7, 4) {};
		\node [style=none] (24) at (9.000001, 4) {};
		\node [style=none] (25) at (6.999999, 7) {};
		\node [style=none] (26) at (9, 7) {};
		\node [style=none] (27) at (10.5, 5.5) {(S2o')};
		\node [style=none] (28) at (2, -0) {};
		\node [style=none] (29) at (1.5, 2) {};
		\node [style=none] (30) at (3.5, 0.9999999) {=};
		\node [style=none] (31) at (5, 2) {};
		\node [style=none] (32) at (5, -0) {};
		\node [style=none] (33) at (6.499999, 0.9999999) {=};
		\node [style=none] (34) at (7.999999, -0) {};
		\node [style=none] (35) at (8.499999, 2) {};
		\node [style=rn] (36) at (1.5, 1) {};
		\node [style=rn] (37) at (0.9999999, -0) {};
		\node [style=rn] (38) at (8.499999, 1) {};
		\node [style=rn] (39) at (9.000001, -0) {};
		\node [style=none] (40) at (10.5, 0.9999999) {(S4o')};
		\node [style=rn] (41) at (-8.5, 1) {};
		\node [style=none] (42) at (-8, -0) {};
		\node [style=none] (43) at (-9, -0) {};
		\node [style=none] (44) at (-8.5, 2) {};
		\node [style=none] (45) at (-6.5, 0.9999999) {=};
		\node [style=none] (46) at (-4, -0) {};
		\node [style=none] (47) at (-4.5, 2) {};
		\node [style=rn] (48) at (-4.5, 1) {};
		\node [style=none] (49) at (-1.5, 1.25) {(S3o')};
		\node [style=none] (50) at (-8, -2) {};
		\node [style=none] (51) at (-8, -5) {};
		\node [style=none] (52) at (-6.5, -3.5) {=};
		\node [style=none] (53) at (-4.5, -2) {};
		\node [style=none] (54) at (-4.5, -5) {};
		\node [style=none] (55) at (-1.5, -3.5) {(S5o')};
		\node [style=none] (56) at (2, -2) {};
		\node [style=none] (57) at (2, -5) {};
		\node [style=none] (58) at (5.999999, -2) {};
		\node [style=none] (59) at (5.999999, -5) {};
		\node [style=rn] (60) at (2, -3) {$\alpha$};
		\node [style=rn] (61) at (2, -4) {$\beta$};
		\node [style=rn] (62) at (5.999999, -3.5) {};
		\node [style=none] (63) at (4, -3.5) {=};
		\node [style=none] (64) at (10.5, -3.5) {(S6o')};
		\node [style=rn] (65) at (-8, -3) {};
		\node [style=rn] (66) at (-8, -4) {};
		\node [style=none] (67) at (7.7, -3.5) {$\alpha+\beta$};
		\node [style=rn] (68) at (-9, 6) {};
		\node [style=rn] (69) at (-9.5, 5) {};
		\node [style=rn] (70) at (-4, 6) {};
		\node [style=rn] (71) at (-3.5, 5) {};
		\node [style=none] (72) at (-5, -0) {};
	\end{pgfonlayer}
	\begin{pgfonlayer}{edgelayer}
		\draw  (1) to (5.center);
		\draw  (8) to (11.center);
		\draw  (17.center) to (14);
		\draw  (14) to (19.center);
		\draw  (15) to (18.center);
		\draw  (22) to (26.center);
		\draw  (21) to (23.center);
		\draw  (22) to (21);
		\draw  (36) to (37);
		\draw  (29.center) to (36);
		\draw  (38) to (39);
		\draw  (35.center) to (38);
		\draw  (31.center) to (32.center);
		\draw  (44.center) to (41);
		\draw  (41) to (42.center);
		\draw  (47.center) to (48);
		\draw  (51.center) to (66);
		\draw [style=simple, bend left=60, looseness=1.50] (65) to (66);
		\draw  (65) to (50.center);
		\draw  (53.center) to (54.center);
		\draw  (56.center) to (60);
		\draw  (60) to (61);
		\draw  (61) to (57.center);
		\draw  (58.center) to (62);
		\draw  (62) to (59.center);
		\draw (2.center) to (0);
		\draw (0) to (1);
		\draw (1) to (4.center);
		\draw (0) to (3.center);
		\draw (12.center) to (7);
		\draw (8) to (10.center);
		\draw (7) to (8);
		\draw (7) to (9.center);
		\draw (14) to (15);
		\draw (15) to (16.center);
		\draw (22) to (25.center);
		\draw (21) to (24.center);
		\draw (41) to (43.center);
		\draw [in=150, out=-120, looseness=1.50] (48) to (46.center);
		\draw (36) to (28.center);
		\draw (38) to (34.center);
		\draw [bend right=60, looseness=1.50] (65) to (66);
		\draw [bend left=60, looseness=1.25] (48) to (72.center);
	\end{pgfonlayer}
\end{tikzpicture}

In particular, these equations, together with (T) and (C), imply: 

\begin{tikzpicture}[scale=0.6]
	\begin{pgfonlayer}{nodelayer}
		\node [style=gn] (0) at (-5, 1.25) {};
		\node [style=gn] (1) at (-5, 0.2499996) {};
		\node [style=gn] (2) at (0, 0.7499997) {};
		\node [style=none] (3) at (-4, -0.7499997) {};
		\node [style=none] (4) at (-6, -0.7499997) {};
		\node [style=none] (5) at (-1, -0.7499997) {};
		\node [style=none] (6) at (1, -0.7499997) {};
		\node [style=none] (7) at (4.000001, -0.7499997) {};
		\node [style=none] (8) at (6, -0.7499997) {};
		\node [style=none] (9) at (2.5, -0) {=};
		\node [style=none] (10) at (-2.5, -0) {=};
		\node [style=none] (11) at (7.500001, -0) {(So')};
	\end{pgfonlayer}
	\begin{pgfonlayer}{edgelayer}
		\draw (0) to (1);
		\draw (1) to (4.center);
		\draw (1) to (3.center);
		\draw [bend right, looseness=1.00] (2) to (5.center);
		\draw [bend left, looseness=1.00] (2) to (6.center);
		\draw [bend left=60, looseness=2.25] (7.center) to (8.center);
	\end{pgfonlayer}
\end{tikzpicture}

therefore we can assume that (So') holds in one direction of the proof. We now add a rule (S') to the new set of circuit equations which is trivially equivalent to (So'): 
\begin{center}
\begin{tikzpicture}[scale=0.55]
	\begin{pgfonlayer}{nodelayer}
		\node [style=Had] (0) at (-2, -0) {N};
		\node [style=gn] (1) at (-4, 2) {};
		\node [style=gn] (2) at (-4, -1) {};
		\node [style=none] (3) at (-2.5, 1) {};
		\node [style=none] (4) at (-3.5, 1) {};
		\node [style=none] (5) at (-1.25, 1) {};
		\node [style=none] (6) at (-0.75, 1) {};
		\node [style=none] (7) at (-3, -1) {};
		\node [style=none] (8) at (-2.5, -1) {};
		\node [style=none] (9) at (-1.5, -1) {};
		\node [style=none] (10) at (-1, -1) {};
		\node [style=none] (11) at (-3, 1) {...};
		\node [style=none] (12) at (-2, -1) {...};
		\node [style=none] (13) at (-4, 3) {};
		\node [style=none] (14) at (0, -0) {=};
		\node [style=none] (15) at (2.5, -1) {};
		\node [style=none] (16) at (2, -1) {...};
		\node [style=Had] (17) at (2, -0) {N};
		\node [style=none] (18) at (1, 1) {};
		\node [style=none] (19) at (3, 1) {};
		\node [style=none] (20) at (3, -1) {};
		\node [style=none] (21) at (2, 1) {...};
		\node [style=none] (22) at (1.5, 1) {};
		\node [style=none] (23) at (1, -1) {};
		\node [style=none] (24) at (2.5, 1) {};
		\node [style=none] (25) at (1.5, -1) {};
		\node [style=none] (26) at (4.5, -0) {(S')};
		\node [style=none] (27) at (-2, 1) {...};
	\end{pgfonlayer}
	\begin{pgfonlayer}{edgelayer}
		\draw (0) to (9.center);
		\draw (0) to (5.center);
		\draw (0) to (6.center);
		\draw (0) to (10.center);
		\draw (0) to (8.center);
		\draw (0) to (7.center);
		\draw (0) to (3.center);
		\draw (4.center) to (0);
		\draw (3.center) to (1);
		\draw (1) to (2);
		\draw (13.center) to (1);
		\draw (17) to (15.center);
		\draw (17) to (24.center);
		\draw (17) to (19.center);
		\draw (17) to (20.center);
		\draw (17) to (25.center);
		\draw (17) to (23.center);
		\draw (17) to (18.center);
		\draw (22.center) to (17);
	\end{pgfonlayer}
\end{tikzpicture} \\
\end{center}
where the N box is an arbitrary ZX network. Adding (So') to the new set of network equations means that we can now assume that (So') holds in both directions of the proof. Note that we only assume that (C) holds in the proof that:$\{$(S1), (S2)$\}$ $\Rightarrow$ $\{$(S1'), (S2'), (S3'), (S4'), (S5'), (S6') $\}$ and not in the other direction.  \\

The equation (S6o') is the same as the equation (S6'). If we assume that (So') and (T) hold, then each of the individual equations (S1o'), (S2o'), (S3o')  and (S5o'), is equivalent to (S1'), (S2'), (S3'), (S4') and (S5') respectively. For example:



shows that (S1o') is equivalent to (S1'). The other four equivalences follow in the same way, by repeatedly using (So'). 

\newpage

\textbf{Lemma A2}: The ZX network equations (B1') and (B2') are equivalent to the (B) rules of the ZX network: 

%

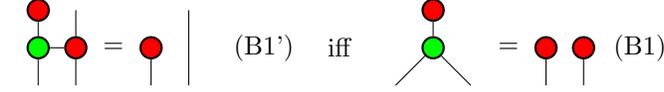
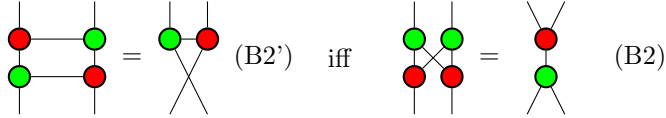
\textbf{Proof:} 
Note that we assume that the rules (T) and (S) hold, which is not a problem since our goal is to prove the equivalence of the whole set of ZX network equations given in Figure \ref{Circax} with the ZX axioms from Figure \ref{ZXRules}. The proof consists of four steps: 

(i) (B1') $\Rightarrow$ (B1):

%
\\[0.4in]

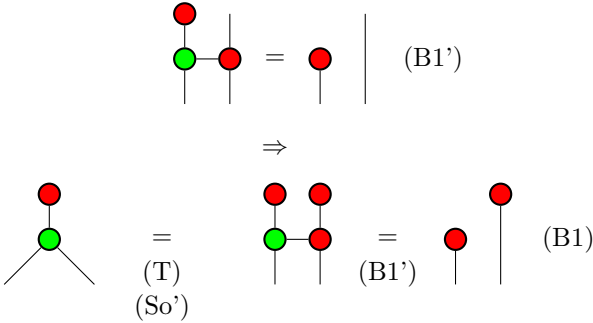
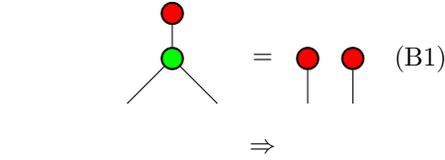
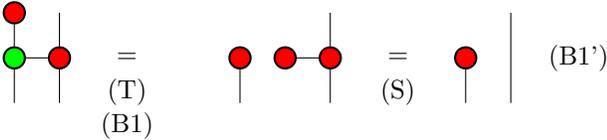
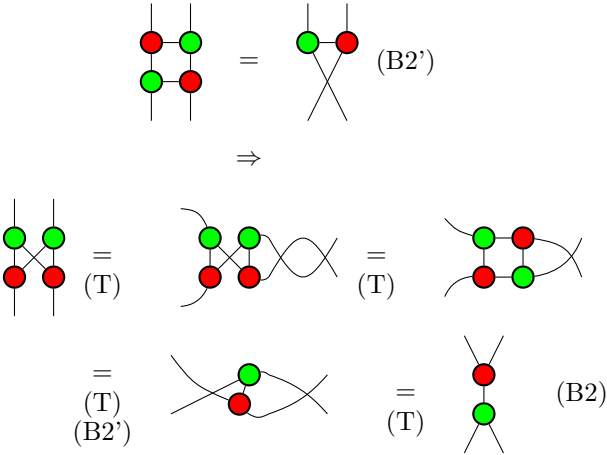
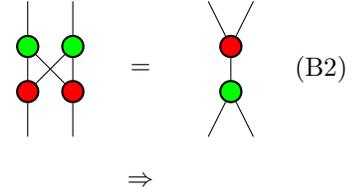
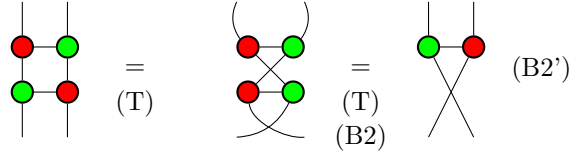
\textbf{Lemma A3}: The ZX network equations (K1') and (K2') are equivalent to the (K) rules of the ZX network: 

%

and (K2') is the same as (K2).

\textbf{Proof:} Once again, we assume that the (S) and (T) rules hold. We show the equivalence in two steps: \\
(i) (K1') ($\Rightarrow$) (K1): 

%
 

\newpage 

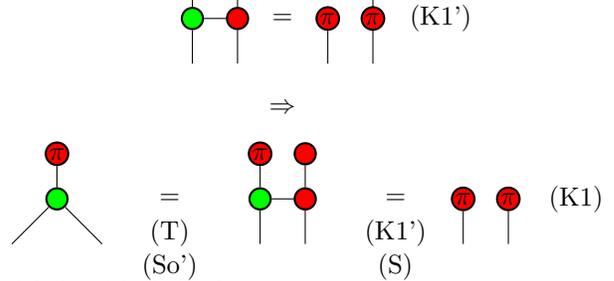
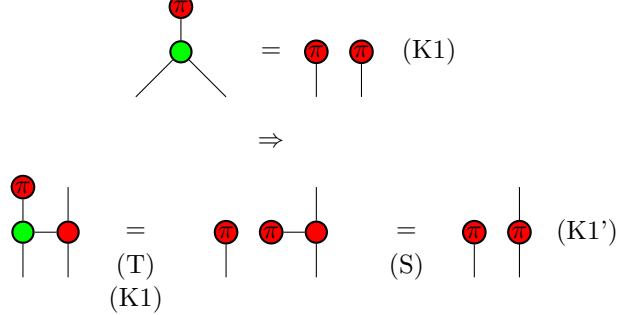
\textbf{Lemma A4}: The ZX network equation (C') is equivalent to the (C) rule of the ZX network: 

%

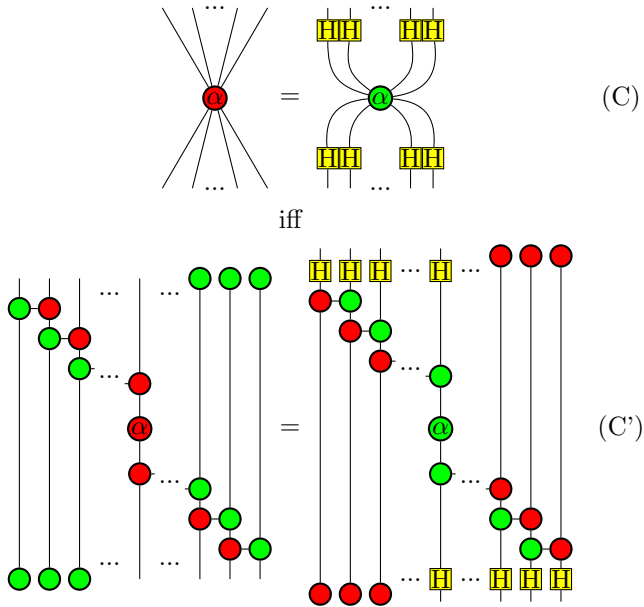
\textbf{Proof:} Again, we assume that the (S) and (T) rules hold. This is not a problem since the proof that $\{$(S1'), (S2'), (S3'), (S4'), (S5'), (S6') $\}$ $\Rightarrow$ $\{$(S1), (S2)$\}$ in Lemma A1 does not assume that (C) holds. The proof of equivalence goes as follows:

%

Therefore, the left and right hand sides of equation (C) are the same as the left and right hand sides respectively of equation (C'), which shows that (C) and (C') are equivalent. Note that both (C) and (C') rules include the case where there are no inputs or no outputs. \\
Note that (H') is the same as (H). Lemmas A1-A4 taken together show that the set of ZX network equations given in Figure \ref{Circax}, are equivalent to the axioms of the ZX network.

\end{document}